\documentstyle[preprint,aps]{revtex}

\begin{document}

\draft

%\preprint{xxx-yy/95}

\title{Axionic Wormholes : \\
More on their Classical and Quantum Aspects }

\author{Hongsu Kim}

\address{Department of Physics\\
Ewha Women's University, Seoul 120-750, KOREA}

\date{April, 1997}

\maketitle

\begin{abstract}
As a system which is known to admit classical wormhole instanton
solutions, Einstein-Kalb-Ramond (KR) antisymmetric tensor theory
is revisited. As an untouched issue, the existence of fermionic
zero modes in the background of classical axionic wormhole
spacetime and its physical implications is addressed. In particular,
in the context of a minisuperspace quantum cosmology model based
on this Einstein-KR antisymmetric tensor theory, ``quantum
wormhole'', defined as a state represented by a solution to the
Wheeler-DeWitt equation satisfying an appropriate wormhole
boundary condition, is discussed. An exact, analytic wave function
for quantum wormholes is actually found. Finally, it is proposed
that the minisuperspace model based on this theory in the presence
of the cosmological constant may serve as an interesting simple
system displaying an overall picture of entire universe's history
from the deep quantum domain all the way to the classical domain.

\end{abstract}


%\narrowtext
%\twocolumn

%%%%%%%%%%%%%%%%%%%%%%%%%%%%%%%%%%%%%%%%%%%%%%%%%%%%%%%%%%%%%%%%%%%%%%%%%%%%%%%%%%%%%%

\centerline {\rm\bf I. Introduction}

At present, in the absence of a complete, consistent theory of quantum
gravity, a systematic formulation of the laws of physics around the
Planck scale seems beyond our scope. Nevertheless, one may still wish
to learn something about these laws by studying possible predictions
from conventional approaches toward the construction of quantum gravity
such as the canonical quantization of general relativity or at least
from its semiclassical approximations. Indeed, some time ago, such
attempts and the associated debates had been circulating in the
theoretical physics society which may be summarized and called as 
``effects of topology change in spacetime on low energy physics'' [1-3].
Known by a more popular name, ``Euclidean wormhole physics'' 
[1-6], these
attempts can be recalled as follows. As pointed out first by Wheeler,
if one identifies the spacetime metric as the relevant gravitational
field subject to the quantization, the topology of spacetime is
expected to fluctuate as well on scales of the order of the Planck length
$l_{p} = M^{-1}_{p}$. And of all types of conceivable spacetime
fluctuations, our major concern is the  ``wormhole configuration''
which is an object that can be loosely defined as the instanton which
is a saddle point of the Euclidean action making dominant contribution
to the topology changing transition amplitude. Then one of the most
crucial effects the wormhole (or more generally, these spacetime
fluctuations) may have on low-energy physics could be the possible
effective loss of quantum coherence [1-3]. For example, one may 
speculate the
situation where ``baby universes'' are pinched off and carry away
information. Then this kind of stereotypical information loss can
lead to an effective loss of quantum coherence as viewed by the
macroscopic observer who cannot measure the quantum state of the baby
universes. At this point, it may be interesting to recall other types
of loss of quantum coherence in semiclassical quantum gravity known
thus far. Namely, from the study of dynamical quantum fields in the 
background of black hole spacetimes, Hawking [9] discovered the 
evaporation of black holes via emitting quanta. He then argued that
this quantum black hole radiation necessarily leads to the ``evolution
of pure states into mixed states'' [10] which obviously signals just 
another type of information loss down to black holes. Although the
calculation involved in the demonstration of this quantum black hole
radiation has been carried out in the context of semiclassical quantum
gravity, the associated quantum incoherence seems generic and thus 
may well survive even the full quantum gravitational treatment.
Therefore these two types of information loses in the wormhole and
in the black hole physics lead us to suspect that dynamics at Planck
scale result in the loss of quantum coherence on general grounds.
In a more careful and concrete analysis, however, these arguments
should be taken with some caution. For example, there is an argument
by Coleman [3] that in the context of ``many universe'' interpretation
(i.e., the third quantization formalism), the quantum incoherence
will not be observed, namely the quantum coherence will be restored.
And there is another argument by Giddings and Strominger [4] which states
that, although at first glance the quantum incoherence is expected
to violate all the conservation laws, what really happens may be that
probably currents associated with local symmetries are exactly
conserved while those associated only with global symmetries are not.
Of all the possible effects of the fluctuations in spacetime topology
on the low energy physics, the most provocative one that immediately
attracted enormous excitement was the advocation initiated by Baum [1]
and by Hawking [1] and then refined later by Coleman [3] that the wormholes
have ultimately an effect of turning all the constants of nature into
``dynamical random variables''. Thus this Baum-Hawking-Coleman (BHC)
mechanism leads to a striking conclusion that the effects of 
(particularly) wormholes introduce into the low energy physics a
fundamental quantum indeterminancy of the values of the constants of
nature which can be thought of as an additional degree of uncertainty
over the usual uncertainty in quantum mechanics. Unlike the issue of
the loss of quantum coherence discussed earlier, however, this 
BHC-mechanism does not simply imply the elimination of the classical
predictability of nature by the effect of quantum gravity. For instance,
the BHC-mechanism actually leads to the prediction that the most
probable value of the fully-renormalized cosmological constant is
zero, indeed in exact agreement with the observation. For the detailed
arguments involved in the BHC-mechanism particularly concerning the
most probable value of the fully-renormalized cosmological constant,
we refer the reader to the literature [1,3]. But it seems fair to mention
that the formulation of wormhole physics particularly the one put
forward by Coleman is not without some inherent flaws. First, logically
the Coleman's wormhole physics formulation had faced some severe
criticisms such as ``sliding of Newton's constant problem'' [7] and
``large wormhole catastrophe'' [8] which seem to be unfortunately quite
generic. However, it seems that the most fundamental and crucial
difficulty associated with Coleman's formulation is the use of saddle
point approximation to the Euclidean path integral for quantum gravity
since, as is well-known, the Euclidean Einstein-Hilbert action is 
not bounded below [11]. \\
Now, then, note that both issues discussed thus far, namely the loss
of quantum coherence and the determination of probability distribution
for constants of nature (which are now random variables), are clearly
based on the assumption that there really are wormhole instantons as saddle
points of the Euclidean action of the theory under consideration.
Therefore unless one can demonstrate that there are large class of
theories comprised of gravity with or without matter which admit
Planck-sized wormhole instantons as solutions to the classical field
equations, the discussion above on interesting effects of wormholes
on low energy physics will lose much of its meaning. Unfortunately,
thus far only a handful of restricted classes of theories are known
to possess classical wormhole instanton solutions and they include ;
Einstein-Kalb-Ramond (KR) antisymmetric tensor theory [4], Einstein-Yang-
Mills theory [5] and Einstein-complex scalar field theory in the presence
of spontaneous symmetry breaking [6]. The classical wormhole instanton
solutions and the semiclassical analysis of their effects on low
energy physics in these theories had been thoroughly studied in the
literature. Here in the present work, we revisit the Einstein-KR
antisymmetric tensor theory of Giddings and Strominger [4] which is a
``classic'' system known to admit classical, Euclidean wormhole
instanton solution. And as was emphasized by Giddings and Strominger [4],
the peculiar feature that renders this theory to possess an Euclidean
wormhole solution is the wrong sign in the Euclidean energy-momentum
tensor when the antisymmetric tensor field strength is represented by
an axion field via duality transformation. One may wonder why anybody
should repeatedly go through a well-studied theory like this one.
Although the ``classical'' wormhole instanton as a solution to the
classical field equations and some of its effects on low energy physics had
been studied extensively, almost no attempt has been made concerning 
the serious study of ``quantum'' wormholes in the same theory.
Besides some important aspects of the classical wormhole physics 
such as the existence of fermion zero modes in the background of 
classical axionic wormhole spacetime and its physical implications
have not been addressed. It is these kinds of untouched but interesting
issues that the present work attempts to deal with. Firstly, the 
investigation of the existence of the fermion zero modes and their
physical implications have been carried out by Hosoya and Ogura[5]  in
wormhole physics in Einstein-Yang-Mills theory. Thus we follow
similar avenue to the one taken there to study the physics associated
with the fermion zero modes in our Einstein-KR antisymmetric tensor
theory. And to do so, we need to introduce interactions between the
KR antisymmetric tensor field and the fermion field possessing both
the general covariance and the local gauge-invariance which has
never been considered thus far. (Note that this interaction should
be distinguished from the derivatively coupled pseudoscalar Goldstone
boson (axion) field - fermion field interactions in effective field
theories.) Secondly, we describe briefly the approach we shall employ
to explore the physics of quantum wormholes in our theory.
And to do so, it seems necessary to distinguish between the definition
of ``classical'' wormholes and that of ``quantum'' wormholes.
In the classical sense, wormholes are Euclidean metrics which are
solutions to the Euclidean classical field equations representing
spacetimes consisting of two asymptotically Euclidean regions joined
by a narrow tube or throat. In the quantum regime, on the other hand,
and particularly in the context of the canonical quantum cosmology,
quantum wormholes may be identified
with a state or an excitation represented by a solution to the
Wheeler-DeWitt equation satisfying a certain boundary condition
describing the wormhole configuration. An widely-accepted such
``wormhole boundary condition'' is the one advocated by Hawking and
Page [12]. And it states that wormhole wave functions are supposed to
behave in such a way that they are damped, say, exponentially for
large 3-geometries ($\sqrt{h} \rightarrow \infty$) and are regular
in some suitable way when the 3-geometry collapses to zero
($\sqrt{h} \rightarrow 0$). Thus we shall construct a minisuperspace
quantum cosmology model possessing SO(4)-symmetry based on the 
Einstein-KR antisymmetry tensor field and attempt to solve associated
Wheeler-DeWitt equation. As we shall see later on, we find an exact,
analytic solution satisfying the ``wormhole boundary condition'' stated
above and identify it with a wormhole wave function, namely a universe
wave function for quantum wormholes. \\
This paper is organized as follows : In sect.2, we recapitulate classical
wormhole instanton solutions and the semiclassical analysis of their
effects on low energy physics in Einstein-KR antisymmetric tensor theory.
In sect.3, we address the issue concerning the fermion zero modes in
the background of classical axionic wormhole spacetime and their
physical implications. Sect.4 will be devoted to the study of quantum 
wormholes in this theory employing the approach described above.
Finally in sect.5, we summarize the results of our study and discuss
their physical implications.
\centerline {\rm\bf II. Axionic Wormhole Instantons Revisited}
\\
We begin by reviewing the classical Einstein-KR antisymmetric tensor theory.\\
Consider a system comprised of an axion (described by a rank-three 
antisymmetric tensor field strength $H_{\mu\nu\lambda}$) coupled to gravity.  
This Einstein-KR antisymmetric tensor (EAT) theory is represented by the 
Euclidean action [4]

\begin{eqnarray}
 I_{EAT} = \int_{M}d^{4}x\sqrt{g}
   [-{{M_{p}^{2}}\over{16\pi}}R + f_{a}^{2}H_{\mu\nu\lambda}H^{\mu\nu\lambda}]
   - \int_{\partial{M}}d^{3}x\sqrt{h}{{M_{p}^{2}}\over{8\pi}}(K-K_{0}) 
\end{eqnarray}
where we added Gibbons-Hawking gravitational boundary term on $\partial M$ with
$h$ being the metric induced on $\partial M$ and $K$ being the trace of
the second fundamental form of $\partial M$.
Here $f_{a}$ is the Peccei-Quinn scale and
$H = dB$ is the field strength tensor of the antisymmetric tensor 
gauge field $B_{\mu\nu}$ of Kalb-Ramond

\begin{eqnarray}
 H = dB.     
\end{eqnarray}
Then since $H = dB$, we have the Bianchi identity $dH = 0$, namely

\begin{eqnarray}
 {\nabla}_{[\rho}H_{\mu\nu\lambda]} = 0. 
\end{eqnarray}
In addition, $H (H_{\mu\nu\lambda})$ is invariant under the gauge 
transformation (since $d^{2}\equiv 0$)

\begin{eqnarray}
 B &&\rightarrow B + d\Lambda \\
 \rm{or}\qquad
 B_{\mu\nu} &&\rightarrow B_{\mu\nu} 
 + (\partial_{\mu}\Lambda_{\nu} - \partial_{\nu}\Lambda_{\mu}).\nonumber
\end{eqnarray}
Now by extremizing the action above with respect to the metric 
$g_{\mu\nu}$ and the Kalb-Ramond antysymmetric tensor field $B_{\mu\nu}$, 
one gets the classical field equations
\begin{eqnarray}
 &&R_{\mu\nu} - {1\over2}g_{\mu\nu}R
 = {{8\pi}\over{M^{2}_{p}}}f_{a}^{2}T_{\mu\nu}  \\
 \rm{with}\quad
 &&T_{\mu\nu} = 6(H_{\mu\alpha\beta}H_{\nu}^{\alpha\beta}
 - {1\over6}g_{\mu\nu}H_{\alpha\beta\gamma}H^{\alpha\beta\gamma}),\nonumber\\
 &&d^{*}H = 0\qquad {\rm or} \qquad \nabla_{\mu}H^{\mu\nu\lambda} = 0, \\
 &&dH = 0 \qquad {\rm or} \qquad \nabla_{[\rho}H_{\mu\nu\lambda]} = 0
\end{eqnarray}
where we included the Bianchi identity in the last line. \\
Now from the classical field equation for the Kalb-Ramond field $d^{*}H = 0$,
we now can define the ``conserved axion current''

\begin{eqnarray}
 j =\ ^{*}H 
\end{eqnarray}
since $dj = d^{*}H = 0$.  Further, we can write, at least, locally 
$j = dA$ (since $d^{2}\equiv 0$) with $A(x)$ denoting the ``axion field''

\begin{eqnarray}
 H &=& ^{*}(dA)   \nonumber \\
{\rm or} ~~~H_{\mu\nu\lambda} &=& \epsilon_{\mu\nu\lambda\beta}
(\partial^{\beta}A)
\end{eqnarray}
only on-shell.
Here, caution must be exercised: if the manifold $M$ is not simply connected, 
the pseudoscalar $A(x)$ may not be globally defined. Now, on-shell, the 
energy-momentum tensor for the Kalb-Ramond field can be expressed in terms of 
the axion field (at least, locally)

\begin{eqnarray}
 T^{\mu\nu} &=& {2\over{\sqrt{g}}}{{{\delta}I_{EAT}}\over{{\delta}g_{\mu\nu}}},
              \nonumber \\
{\rm namely,}
 ~~~T_{\mu\nu}(A) &=& -12 [(\nabla_{\mu}A)(\nabla_{\nu}A) 
                      - {1\over2}g_{\mu\nu}
                     (\nabla_{\lambda}A)(\nabla^{\lambda}A)]. 
\end{eqnarray}
As was first pointed out by Giddings and Strominger [4], the energy-momentum
tensor expressed in terms of the axion field $A(x)$ has wrong sign when
compared with that of ordinary, minimally-coupled scalar field. Thus, the
Euclidean behavior of the axion field (associated with the KR antisymmetric
tensor field) coupled to gravity is radically different from that of an
ordinary scalar field and this is essentially responsible for the fact
that wormhole instantons do exist in this Einstein-KR antisymmetric tensor
theory. \\
Now, we look for a Euclidean $SO(4)$-symmetric wormhole solution which is an 
instanton describing the nucleation of a Planck-sized baby 
(spatially-closed; $k = +1$) FRW universe.  To this end, we begin by taking 
$SO(4)$-symmetric ans\H atz for the Euclidean metric and Kalb-Ramond 
antisymmetric tensor field as

\begin{eqnarray}
 ds^{2} &=& g_{\mu\nu}dx^{\mu}dx^{\nu} =
            \delta_{AB} e^{A}{\otimes}e^{B} \nonumber \\
        &=& N^{2}(\tau)d{\tau}^{2} + a^{2}(\tau)d{\Omega}_{3}^{2}, \\
 H      &=& h(\tau)\epsilon \qquad {\rm or}\qquad 
            H_{\mu\nu\lambda} = h(\tau)\epsilon_{\mu\nu\lambda} 
            \qquad (\mu,\nu,\lambda \not= \tau) \nonumber
\end{eqnarray}
where $e^{A}$ $(A,B = 0,1,2,3)$ are non-coordinate basis 1-forms
\begin{eqnarray}
e^{A} = \{e^{0} = Nd\tau, ~~e^{a} = a\sigma^{a}\},
\end{eqnarray}
$N(\tau)$ and $a(\tau)$ are lapse function and scale factor respectively
and $d{\Omega}^{2}_{3} = \sigma^{a}{\otimes}\sigma^{a}$
denotes the line element on 3-sphere $S^{3}$ with $\{\sigma^{a}\}$
$(a = 1,2,3)$ forming a basis on the $S^{3}$ and
$\epsilon = {1\over3!}\epsilon_{\mu\nu\lambda}
                dx^{\mu}{\wedge}dx^{\nu}{\wedge}dx^{\lambda} 
              = \sqrt{h}d^{3}x$ 
is the volume 3-form normalized so that 

\begin{eqnarray}
 \int_{S^{3}}\epsilon = \int_{S^{3}}d^{3}x\sqrt{h} = 2{\pi}^{2}a^{3}(\tau).
\end{eqnarray}
Then for later use, we note that in terms of this SO(4)-symmetric ans\H atz,
the energy-momentum tensor for the KR antisymmetric tensor field becomes

\begin{eqnarray}
 T_{\mu\nu} &=&  6(H_{\mu\alpha\beta}H_{\nu}^{\alpha\beta} 
 - {1\over6}g_{\mu\nu}H_{\alpha\beta\gamma}H^{\alpha\beta\gamma})\nonumber\\
            &=&  6h^{2}(\tau)(2h_{\mu\nu} - g_{\mu\nu})          \\
 {\rm with}\qquad &&h_{\tau\tau} = h_{\mu\tau} = h_{\tau\nu} = 0.
                                               \nonumber 
\end{eqnarray}
We now have to solve the ``coupled'' Einstein-antisymmetric tensor field 
equations.  Fortunately, however, the antisymmetric tensor sector of field 
equations, namely the Euler-Lagrange's equation of motion and the Bianchi 
identity are satisfied,
$ d^{*}H = 0  = dH $,
if we set [4]
\begin{eqnarray}
 h(\tau) = {n\over{f_{a}^{2}a^{3}(\tau)}} 
\end{eqnarray}
so that
\begin{eqnarray}
 \int_{S^{3}}H = \int_{S^{3}}h(\tau)\epsilon 
 = {{2{\pi}^{2}n}\over{f_{a}^{2}}}  
\end{eqnarray}
where $f_{a}$ is, as introduced, the Peccei-Quinn scale and $n$ is a
free parameter which, in string theory, is quantized.
In fact, once we set $H_{\mu\nu\lambda} = h(\tau)\epsilon_{\mu\nu\lambda}$, 
it automatically satisfies the Euler-Lagrange's equation of motion 
$\nabla_{\mu}H^{\mu\nu\lambda} = 0$.
Note here that in contrast to the original formulation by Giddings and
Strominger [4] where they obtained this SO(4)-symmetric expression for the 
KR antisymmetric tensor field strength as a solution to the classical
field equation and the Bianchi identity, here we stress that it can be
``derived'' simply from the definition $H = dB$ and the Bianchi identity
$dH = 0$ {\it without} imposing on-shell condition. 
To see this briefly, using the SO(4)-symmetric ans\H atz
$H_{\mu\nu\lambda} = h(\tau)\epsilon_{\mu\nu\lambda}$, start with
\begin{eqnarray}
H &=& {1\over 3!}H_{\mu\nu\lambda}dx^{\mu}\wedge dx^{\nu}\wedge
      dx^{\lambda} \nonumber \\
  &=& {1\over 3!}h(\tau)\epsilon_{abc}(e^{a}\wedge e^{b}\wedge e^{c})
      \nonumber
\end{eqnarray}
where $a,b,c = 1,2,3$ since $\mu, \nu, \lambda \not= \tau$.
Then consider
\begin{eqnarray}
dH = {1\over 3!}[h'(\tau)\epsilon_{abc}(d\tau \wedge e^{a}\wedge
     e^{b}\wedge e^{c}) + h(\tau)\epsilon_{abc}d(e^{a}\wedge
     e^{b}\wedge e^{c})]. \nonumber
\end{eqnarray}
Since $\{\sigma^{a}\}$, which form a basis on the three-sphere
$S^{3}$, satisfy the SU(2) ``Maurer-Cartan'' structure equation
$d\sigma^{a} = {1\over 2}\epsilon^{abc} \sigma^{b}\wedge \sigma^{c}$,
we have, using eq.(12),
\begin{eqnarray}
dH = {1\over 3!}[({h'\over N} + 3{h\over N}{a'\over a})\epsilon_{abc}
     (e^{0}\wedge e^{a}\wedge e^{b}\wedge e^{c})]. \nonumber
\end{eqnarray}
Finally, imposing the Bianchi identity $dH = 0$ (since $H = dB$) yields
\begin{eqnarray}
h' + 3({a'\over a})h = 0 \nonumber
\end{eqnarray}
of which the solution takes the form given in eq.(15),
\begin{eqnarray}
h(\tau) = {n\over f^2_{a}a^3(\tau)}. \nonumber
\end{eqnarray}
Therefore, this
SO(4)-symmetric ans\H atz for the KR antisymmetric field strength in 
eq.(15) remains perfectly valid even off-shell as well as on-shell.
In other words, this expression for $H_{\mu\nu\lambda}$ can be used
for both classical and quantum treatments that we shall discuss later on
in the section of quantum axionic wormholes. \\
Now what remains is to solve the Einstein field equations which are no longer 
coupled equations and reduce to non-linear equations of the scale factor 
$a(\tau)$ alone.  Besides, we only need to consider the time-time component 
of the Einstein equations since the rest of the equations are implied by the 
Bianchi identity (i.e., energy-momentum conservation).  
Thus from eqs. (5), (11) and (14),
consider the $\tau\tau$-component of the Einstein equations

\begin{eqnarray}
 {{3M_{p}^{2}}\over{16\pi}}[({a'\over{a}})^{2} - {{N^{2}}\over{a^{2}}}]
 = -{{3n^{2}}\over{f_{a}^{2}}}{{N^{2}}\over{a^{6}}}. \nonumber
\end{eqnarray}
where the ``prime'' denotes the derivative with respect to the Euclidean 
time $\tau$. \\
(1) With the gauge-fixing $N(\tau) = 1$ :

\begin{eqnarray}
 ({a'\over{a}})^{2} - {1\over{a^{2}}} = -{r^{4}\over{a^{6}}}\qquad
 \Bigl(r^{2} = {{4\sqrt{\pi}n}\over{M_{p}f_{a}}} = a^{2}(\tau =0)\Bigr).
\end{eqnarray}
Since we set $r^{2}$ as the value of $a^{2}$ at $\tau = 0$, 
i.e., $a^{2}(\tau = 0) = r^{2}$, it can be integrated to yield

\begin{eqnarray}
 \tau &=& \int^{\tau}_{0}d{\tau}' = \int^{a(\tau)}_{a(0)=r^{2}}
 {{a^{2}da}\over{\sqrt{(a^{2} + r^{2})(a^{2} - r^{2})}}} \\
 &=& {r\over{\sqrt{2}}}F[cos^{-1}({r\over{a}}),{1\over{\sqrt{2}}}]
  -  \sqrt{2}rE[cos^{-1}({r\over{a}}),{1\over{\sqrt{2}}}]
  +  {1\over{a}}\sqrt{a^{4} - r^{4}} \nonumber
\end{eqnarray}
where $F$ and $E$ are elliptic integrals of the first and second
kinds, respectively.
This Euclidean wormhole instanton solution is characterized by one free 
parameter, $n$, which, as mentioned, is quantized in string theory.  
Note the ``asymptotic behavior'' of this Euclidean wormhole solution

\begin{eqnarray}
 a(\tau)\rightarrow\tau, \qquad {\rm as}\quad 
 \tau\rightarrow\pm\infty.   \nonumber
\end{eqnarray}
The wormhole instanton solution we obtained is drawn in Fig. 1.  
Since $a^{2}(\tau)\rightarrow\tau^{2} $ as $\tau\rightarrow\pm\infty$, 
there are two asymptotically Euclidean regions.  They are joined by 
a ``throat'' whose cross sections are $S^{3}$'s.  The axion current $^{*}j$ 
has total integrated flux $n/f_{a}^{2}$ through the throat.  As it stands, 
it is difficult to ascribe a physical interpretation to this instanton 
configuration because of the two asymptotic regions. 
(However, it might represent communication between two different universes).  
The situation can be improved by slicing the wormhole instanton in half 
through the minimal surface of the throat which is drawn in Fig. 2.  
It represents tunnelling from an initial hypersurface $\Sigma_{i}$ with 
topology $R^{3}$ to a final hypersurface $\Sigma_{f}$ with topology 
$R^{3}{\oplus}S^{3}$.  It describes nucleation of a baby (spatially-closed) 
FRW-universe created at its moment of time symmetry.  Note that, in order for 
this instanton solution in Fig. 2 to describe a reasonable tunnelling process,
 it is crucial that the fields involved (i.e., the metric field $a(\tau)$ 
and the Kalb-Ramond field strength 
$H_{\mu\nu\lambda} = h(\tau)\epsilon_{\mu\nu\lambda}
~(\mu,\nu,\lambda \not= \tau)$) should take appropriate values.  Namely, 
it must be true that the fields and their first time derivatives on 
$\Sigma_{i}$ and $\Sigma_{f}$ are all real when analytically continued back 
to the Lorentzian spacetime.  This is obvious for the $R^{3}$ part of 
$\Sigma_{i}$ and $\Sigma_{f}$.  On the $S^{3}$ portion, the time derivative 
of the metric vanishes because it is a minimal surface.  The time components 
of $H_{\mu\nu\lambda} = h(\tau)\epsilon_{\mu\nu\lambda}$ vanish because it is 
a 3-form tangent to the spacelike hypersurface.   Thus the instanton obtained 
above does obey exactly the right boundary conditions for the description 
of the tunnelling $R^{3}\rightarrow R^{3}\oplus S^{3}$.  An additional 
important feature which characterizes this instanton is the axion current 
through the throat of the wormhole ($R\otimes S^{3}$) and the axion charge 
on the non-contractable 3-spheres ($S^{3}$).  Axion current through 
the wormhole throat is from eq.(8),

\begin{eqnarray}
 ^{*}j = H = ^{*}(dA) 
\end{eqnarray}
which is a 3-form and is conserved owing to Bianchi identity
$ d^{*}j = dH = 0$.
The axion current flux through the wormhole throat is 

\begin{eqnarray}
 \int_{S^{3}}\ ^{*}j =\int_{S^{3}}H =\int_{S^{3}}h(\tau)\epsilon
  = \int_{S^{3}}\epsilon{n\over{f_{a}^{2}a^{3}(\tau)}} 
  = 2{\pi}^{2}({n\over{f_{a}^{2}}}).
\end{eqnarray}
Axion charge on the cross section of the wormhole throat is 

\begin{eqnarray}
 q = f_{a}^{2}\int_{S^{3}(\tau=0)}\ ^{*}j = f_{a}^{2}\int_{S^{3}(\tau=0)}H
   = 2{\pi}^{2}n. 
\end{eqnarray}
(In string theory, global anomalies in the string sigma model lead to 
quantization of $n$ in this axion charge.)  The observer on $R^{3}$ will 
measure a change of ${1\over{2\pi^2}}{\triangle}q = (-n)$ in the axion charge,
 since the baby universe pinches off $n$-units of axion charge.  This, 
of course, would be rather puzzling to the observer on $R^{3}$, who cannot 
observe the charge on the baby universe, since he or she may believe that 
axion 
charge is conserved due to the (unbroken since $f_{a} < M_{p}$) Peccei-Quinn 
symmetry of the action.  This effective charge non-conservation can be 
understood as a result of the breakdown of quantum coherence due to 
information loss to baby universes (in the similar spirit to information loss 
in black hole evaporatioin by Hawking effect).  Next, we evaluate the 
(wormhole) instanton actioin $I_{EAT}$(instanton), namely the minimum 
Euclidean action of the instanton configuration which makes dominant 
contribution to the tunnelling amplitude.  Namely, into the Euclidean action 
of this Einstein-antisymmetric tensor theory given earlier

\begin{eqnarray}
 I_{EAT} = \int_{M}d^{4}x\sqrt{g}
   [-{{M_{p}^{2}}\over{16\pi}}R + f_{a}^{2}H^{2}] \nonumber 
\end{eqnarray}
we substitute the Einstein field equation (its trace) satisfied by the 
wormhole instanton solution,

\begin{eqnarray}
 R = -{{16\pi }\over {M_{p}^{2}}}f_{a}^{2}H^{2}
\end{eqnarray}
to obtain $\Bigl({\rm using}\quad H 
= h(\tau)\epsilon = {n\over{f_{a}^{2}a^{3}(\tau)}}\epsilon\quad {\rm and}\quad
\int_{M}d^{4}x\sqrt{g}=\int^{\infty}_{0}d{\tau}\int_{S^{3}}\epsilon 
= 2{\pi}^{2}\int^{\infty}_{0}d{\tau}a^{3}(\tau)\Bigr)$

\begin{eqnarray}
 I_{EAT}({\rm instanton})
 &=& 2f_{a}^{2}\int_{M}d^{4}x\sqrt{g}H^{2}
 = {{24{\pi}^{2}n^{2}}\over{f_{a}^{2}}}
   \int^{\infty}_{0}d{\tau}{1\over{a^{3}(\tau)}} \\
 &=& {{6{\pi}^{3}n^{2}}\over{f_{a}^{2}r^{2}}} 
 = {3{\pi}^{2}\over8}r^{2}M_{p}^{2}
 = {3{\pi}^{2}\over8}({r\over{l_{p}}})^{2}
 = {{3{\pi}^{5/2}nM_{p}}\over{2f_{a}}}. \nonumber
\end{eqnarray}
Recall that the amplitude for the tunnelling from $R^{3}$ to 
$R^{3}{\oplus}S^{3}$, namely the ``baby universe nucleation rate'' is 
proportional to 

\begin{eqnarray}
 e^{-I_{EAT}({\rm instanton})}
 = \exp{[-{{3{\pi}^{2}}\over8}({r\over{l_{p}}})^{2}]}
\end{eqnarray}
where $l_{p} = M_{p}^{-1}$ is the Planck length.  Thus, first of all, we see 
that fortunately the rate for the nucleation of the 
baby universes (or wormholes) with 
size larger than the Planck length $(r>l_{p})$ is highly suppressed.
Also, we can see that a typical baby universe (or wormhole) 
will have a radius of order $r{\sim}\sqrt{{8\over3{\pi}^{2}}}l_{p}$.  
Another quantity which is important in determining the effects of wormhole 
instantons on the low energy physics is the axion charge carried away by 
the baby universe.  This has the typical value of 
$n\sim({2f_{a}\over3{\pi}^{5/2}M_{p}}).$  Finally, when the wormhole 
instantons or equivalently nucleated baby universes are widely separated, 
the ``dilute instanton gas approximation'', in which the interactions 
between instantons can be ignored, can be valid to be used.
For the sake of completeness, next we also consider the physics of classical
wormhole solution resulting from an alternative gauge choice for the
lapse function $N(\tau)$. \\
(2) With the conformal-time gauge fixing $N(\tau) = a(\tau)$ :
\begin{eqnarray}
 ({{a}'\over{a}})^{2} - 1 = -{r^{4}\over{a}^{4}}. \qquad
 \Bigl(r^{2}&\equiv&{4\sqrt{\pi}n\over{M_{p}f_{a}}} 
 = a^{2}(\tau=0)\Bigr) \\
\end{eqnarray}
which yields, upon integration,

\begin{eqnarray}
 a(\tau) &=& r[cosh(2\tau)]^{1/2}, \\
 h(\tau) &=& {n\over{f_{a}^{2}a^{3}(\tau)}}
          =  {n\over{f_{a}^{2}r^{3}}}[cosh(2\tau)]^{-3/2}. \nonumber 
\end{eqnarray}
Note the asymptotic behavior of this Euclidean wormhole solution

\begin{eqnarray}
 a(\tau) \rightarrow &&{r\over{\sqrt{2}}}e^{\tau}\qquad
 ({\rm as}\quad\tau\rightarrow\infty),\nonumber\\
 \rightarrow &&{r\over{\sqrt{2}}}e^{-\tau}\qquad
 ({\rm as}\quad\tau\rightarrow-\infty).\nonumber
\end{eqnarray}
Thus again, this wormhole solution represents a configuration in which two 
asymptotically-Euclidean regions are connected by a wormhole with throat 
(or neck) whose cross sections are $S^{3}$'s with minimum radius 
$a(\tau = 0) = r = ({4\sqrt{\pi}n\over{M_{p}f_{a}}})^{1/2}$.  
Next, as before, we evaluate the (wormhole) instanton action 
$I_{EAT}$(instanton), namely the minimum Euclidean action of the instanton 
configuration which makes dominant contribution to the tunnelling amplitude, 
i.e., baby universe nucleation rate.  And this amounts to substituting the 
Einstein field equation (its trace) satisfied by the wormhole instanton 
solution into the Euclidean Einstein-antisymmetric tensor theory action  
as we did before.
Now for the case at hand where we take the ``conformal time gauge'' 
for the lapse function, $N(\tau) = a(\tau)$, 
using $H = h(\tau)\epsilon = {n\over{f_{a}^{2}a^{3}(\tau)}}\epsilon$ 
and $\int_{M}d^{4}x\sqrt{g} 
= \int^{\infty}_{0}d{\tau}N(\tau)\int_{S^{3}}\epsilon 
= 2{\pi}^{2}\int^{\infty}_{0}d{\tau}N(\tau)a^{3}(\tau)$, we obtain 

\begin{eqnarray}
 I_{EAT}({\rm instanton})
 &=& 2f_{a}^{2}\int_{M}d^{4}x\sqrt{g}H^{2}
  = {24{\pi}^{2}n^{2}\over{f_{a}^{2}}}\int^{\infty}_{0}
    d{\tau}N(\tau)a^{3}(\tau){1\over{a^{6}(\tau)}} \\
 &=& {24{\pi}^{2}n^{2}\over{f_{a}^{2}}}\int^{\infty}_{0}
    {d{\tau}\over{a^{2}(\tau)}}
  = {6{\pi}^{3}n^{2}\over{f_{a}^{2}r^{2}}}
  = {3{\pi}^{2}\over8}({r\over{l_{p}}})^{2}
  = {3{\pi}^{5/2}nM_{p}\over{2f_{a}}}. \nonumber
\end{eqnarray}
Notice here that although the form of spacetime metric $a(\tau)$ and the 
Kalb-Ramond 
field strength $h(\tau)$ solution are different for two different gauge 
choices $N(\tau) = 1$ and $N(\tau) = a(\tau)$, the Euclidean instanton action 
evaluated at these wormhole instanton configurations remains the same 
indicating the same tunnelling process.  Namely, since the quantity 
$\sim \exp{[-I_{EAT}({\rm instanton})]}$ represents the semi-classical 
approximation to the baby universe nucleation rate, it is a physical 
observable which should have manifest gauge-invariance.  The relevant gauge 
freedom for the case at hand is the arbitrariness in choosing the lapse 
$N(\tau)$ and we have just confirmed this gauge-invariance.  Indeed, this 
gauge-invariance of the instanton action $I_{EAT}$(instanton) and hence 
that of 
the semi-classical baby universe nucleation rate can be generally displayed 
as follows;  from the general form of the time-time component of Einstein 
equations 

\begin{eqnarray}
 ({a'\over{a}})^{2} - {N^{2}\over{a^{2}}} = -r^{4}{N^{2}\over{a^{6}}},
\end{eqnarray}
we get

\begin{eqnarray}
 d\tau = {{a^{2}da}\over{N\sqrt{a^{4} - r^{4}}}}.
\end{eqnarray}
Meanwhile, the general form of the Euclidean instanton action is given by 
(using $\int_{M}d^{4}x\sqrt{g} = 2{\pi}^{2}\int^{\infty}_{0}d{\tau}Na^{3}$)

\begin{eqnarray}
 I_{EAT}({\rm instanton})
 &=& 2f^{2}_{a}\int_{M}d^{4}x\sqrt{g}H^{2} \\
 &=& {24{\pi}^{2}n^{2}\over{f_{a}^{2}}}\int^{\infty}_{0}{N\over{a^{3}}}d\tau.
     \nonumber
\end{eqnarray}
Thus by using eq.(30), we finally obtain

\begin{eqnarray}
 I_{EAT}({\rm instanton})
 = {24{\pi}^{2}n^{2}\over{f_{a}^{2}}}\int^{\infty}_{r}
   {da\over{a\sqrt{a^{4} - r^{4}}}}
 = {6{\pi}^{3}n^{2}\over{f_{a}^{2}r^{2}}}.
\end{eqnarray}
We now end this section with some comments.  As was noted by Rey [4], 
the axionic 
instantons (or wormholes) are characterized by their axion charges $n$ which 
are necessary for their stability.  Therefore, if we call the axionic 
instanton with the axionic charge $+n$ as an instanton, then we may identify 
that with the axionic charge $-n$ as its ``anti-instanton''.  Then denoting 
an instanton with axion charge $+n$ and an anti-instanton 
with $-n$ by $I_{n}$ and $I_{-n}$ respectively, we may speculate 
the pair-annihilation-type process such as 

\begin{eqnarray}
 I_{n} + I_{-n} \quad\leftrightarrow\quad({\rm flat\ spacetime}).\nonumber
\end{eqnarray}
Indeed, the possibility of this process is supported by the fact that for 
cases when the dilute gas approximation is valid, 
$I_{EAT}$(instanton) + $I_{EAT}$(anti-instanton) = $Cn + C(-n) = 0$ 
(where $C\equiv{3{\pi}^{5/2}M_{p}\over{2f_{a}}}$), namely the total action 
of the instanton-anti-instanton system is zero which is the action of the 
vacuum, i.e., flat spacetime.  And this statement may remain true for any 
other theory involving wormhole solution which is stabilized 
by global charges of the underlying physics. Next, if we carefully 
distinguish between Fig.1 and 2 such that Fig.1 depicts ``wormhole'' 
configuration and Fig.2, ``instanton'' configuration, the wormhole 
configuration can be identified with a ''bounce'' solution 
of $R^{3}\rightarrow R^{3}{\oplus}S^{3}\rightarrow R^{3}$. Then a wormhole 
with axionic charge $+n$ is the double of two axionic instantons 
with charge $+n$ and $-n$ respectively or equivalently of two 
oppositely-oriented instantons sewed together along the (Euclidean) spacelike 
boundaries $S^{3}$'s.  Finally, taking our $SO(4)$-symmetric homogeneous 
and isotropic wormhole and axion field solution as the ground state 
(i.e., maximally-symmetric) solution, one may wish to look for exited 
wormhole and axion field solutions with, say, slightly broken 
$SO(4)$-symmetry.  For example, one may try with the wormhole solution 
ans\H atz being given by the Bianchi type-IX metric [13] which is still 
homogeneous but not exactly isotropic.  

\centerline {\rm\bf III. The fermion zero modes and their effects 
on low-energy physics}

As have been pointed out first by Hosoya and Ogura [5], but in a different 
context where they studied wormhole instantons in Einstein-Yang-Mills 
theory, 
if there exist fermionic zero modes in a given background of wormhole 
spacetime, they may have profound effects on low-energy physics presumably 
in a similar manner the instanton configurations in non-abelian gauge 
theories do.  To name one, one may expect that chirality-changing fermion 
propagation in the background of a wormhole instanton could arise.  In order 
to see if this kind of intriguing possibility can actually happen 
in our case of axionic wormhole instanton, we first attempt to investigate 
the existence of normalizable fermion zero modes in the background of axionic 
wormhole instantons.  Thus far, we have considered the theory of free KR 
(classical) antisymmeric tensor field $B_{\mu\nu}$ coupled only to gravity.  
In order to explore the dynamics of fermion field in the background 
of classical wormhole instanton solutions in Einstein-antisymmetric tensor 
theory,  we need to know the fermion-KR antisymmetric tensor field 
interaction as well as the fermion-gravity minimal coupling.  
To our knowledge, however, the fermion-KR antisymmetric tensor gauge field 
interaction ({\it not} the Yukawa-type axion-axial fermion current 
interactions $A(x)\bar{\Psi}\gamma_{5}\Psi$ 
or $\partial_{\mu}A(x)\bar{\Psi}\gamma^{\mu}\gamma_{5}\Psi$ 
in effective Lagrangians of extended standard model based on the gauge group 
$SU(3)_{C}{\times}SU(2)_{L}{\times}U(1)_{Y}{\times}U(1)_{PQ}$ 
with $U(1)_{PQ}$ being the anomalous Peccei-Quinn symmetry group) is not 
known nor has been seriously considered yet.  Since the pure KR antisymmetric 
tensor gauge theory possesses a gauge invariance based on abelian 
$\Bigl(U(1)\Bigr)$ gauge group as mentioned earlier, one can construct 
a fermion-KR antisymmetric tensor gauge field interaction Lagrangian 
which has a manifest local $U(1)$ gauge-invariance.  
Here in the present work, we propose, as one such attempt, a theory 
of massless fermion-KR antisymmetric tensor field system involving 
both the local tensor $U(1)_{V}$ and axial tensor $U(1)_{A}$ gauge-invariant 
couplings described by the action in flat Minkowski spacetime

\begin{eqnarray}
 S_{KR-F} = {\int}d^{4}x[-f_{a}^{2}tr(H_{\mu\nu\lambda}H^{\mu\nu\lambda}) 
          + \bar{\Psi}i\gamma^{\mu}D_{\mu}\Psi]
\end{eqnarray}  
with the gauge-covriant derivative being given by 

\begin{eqnarray}
 \gamma^{\mu}D_{\mu}
 &\equiv&(\gamma^{\mu}\partial_{\mu} - \sigma^{\mu\nu}B_{\mu\nu}^{V} 
      + \sigma^{\mu\nu}\gamma_{5}B_{\mu\nu}^{A})\nonumber\\
 &=& (\gamma^{\mu}\partial_{\mu} - i\gamma^{\mu}\gamma^{\nu}B_{\mu\nu}^{V} 
      + i\gamma^{\mu}\gamma^{\nu}\gamma_{5}B_{\mu\nu}^{A}) \\
 &=& \gamma^{\mu}(\partial_{\mu} - i\gamma^{\nu}B_{\mu\nu}^{V} 
      + i\gamma^{\nu}\gamma_{5}B_{\mu\nu}^{A})\nonumber
\end{eqnarray}
where $\sigma^{\mu\nu} = {i\over2}[\gamma^{\mu},\gamma^{\nu}]$ 
and we let $B^{VA}_{\mu\nu}\rightarrow B^{VA}_{\mu\nu}I$ (with $I$ being the 
$4\times4$ identity matrix which explains ``$tr$'' in the KR field 
term in the action above. It is straightforward to check that this action 
is invariant under the local tensor $U(1)_{V}$ and axial tensor $U(1)_{A}$ 
transformations given by

\begin{eqnarray} 
 B^{VA}_{\mu\nu}\rightarrow B^{VA}_{\mu\nu} 
 + (\partial_{\mu}\Lambda^{VA}_{\nu} - \partial_{\nu}\Lambda^{VA}_{\mu})
\end{eqnarray}
where

\begin{eqnarray}
 \Lambda^{\mu}_{VA}(x) = {1\over2(n-1)}\gamma^{\mu}\theta_{VA}(x)\qquad
 {\rm (say,\ in}\ n-{\rm dim.)}
\end{eqnarray}
along with

\begin{eqnarray}
 \Psi\rightarrow e^{i\theta_{V}(x)}\Psi\qquad,\qquad
 \bar{\Psi}\rightarrow \bar{\Psi}e^{-i\theta_{V}(x)}
\end{eqnarray}
for local $U(1)_{V}$ transformation and 

\begin{eqnarray}
 \Psi\rightarrow e^{i\gamma_{5}\theta_{A}(x)}\Psi\qquad,\qquad
 \bar{\Psi}\rightarrow \bar{\Psi}e^{i\gamma_{5}\theta_{A}(x)}
\end{eqnarray}
for local $U(1)_{A}$ transformation.
The guideline for our choice of the fermion-KR antisymmetric tensor gauge 
field interaction given above is as follows; 
certainly we need minimal coupling in which KR field 
itself $B_{\mu\nu}(x)$, not its field strength $H_{\mu\nu\lambda}(x)$, 
is supposed to be present as is obvious from our experience with ordinary 
vector gauge field theories.  Next, since the KR tensor gauge field 
$B_{\mu\nu}^{V}(x)$ is antisymmetric under interchange of its indices, 
antisymmetric fermion tensor current of rank-2, 
$J^{\mu\nu} = \bar{\Psi}\sigma^{\mu\nu}\Psi$, which appears to be the only 
choice available, should couple to it, i.e., 
$J^{\mu\nu}B_{\mu\nu}^{V} = \bar{\Psi}\sigma^{\mu\nu}\Psi B_{\mu\nu}^{V}.$ 
And a similar argument applies to our choice of KR axial tensor gauge field 
$B_{\mu\nu}^{A}$-fermion axial tensor current coupling term, 
$J^{\mu\nu}_{5}B^{A}_{\mu\nu} 
= \bar{\Psi}\sigma^{\mu\nu}\gamma_{5}{\Psi}B^{A}_{\mu\nu}$.  
Now, since the examination of the dynamical fermion fields in the background 
of classical KR antisymmetric tensor field and curved spacetime 
(i.e., wormhole geometry) is of our present interest in this work, next we 
consider the case when the gravity is turned on (but just as a ``background'' 
field) with the KR antisymmetric tensor field freezing again 
as a non-dynamical degree.  Then the theory of a dynamical fermion field 
in the background of KR and gravitational field would naturally be described 
by the action

\begin{eqnarray}
 S_{F} &=& {\int}d^{4}x ~e ~{i\over2}
 [\bar{\Psi}\gamma^{\mu}\overrightarrow{\nabla}_{\mu}\Psi 
 - \bar{\Psi}\gamma^{\mu}\overleftarrow{\nabla}_{\mu}\Psi] \\
 &=& {\int}d^{4}x ~e ~{i\over2}
 [\bar{\Psi}\gamma^{A}e^{\mu}_{A}\overrightarrow{\nabla}_{\mu}\Psi 
 - \bar{\Psi}\gamma^{A}e^{\mu}_{A}\overleftarrow{\nabla}_{\mu}\Psi]\nonumber
\end{eqnarray}
where the covariant derivative now generalizes to 

\begin{eqnarray}
 \gamma^{\mu}\nabla_{\mu}
 &\equiv&[\gamma^{\mu}(\partial_{\mu} - {i\over4}\omega_{\mu}^{AB}\sigma_{AB}) 
        - \sigma^{\mu\nu}B_{\mu\nu}] \\
 &=& \gamma^{C}e^{\mu}_{C}
     [\partial_{\mu} - {i\over4}\omega^{AB}_{\mu}\sigma_{AB} 
        - i\gamma^{B}e^{\nu}_{B}B_{\mu\nu}].\nonumber
\end{eqnarray}
(Here we consider only the KR tensor field - fermion tensor current coupling, 
$J^{\mu\nu}B_{\mu\nu}^{V}$, which is of usual relevance.)  
Then the corresponding Dirac equations for massless fermion field are given by

\begin{eqnarray}
 &&\gamma^{C}e^{\mu}_{C}
 [\overrightarrow{\partial}_{\mu} 
 - {i\over4}\omega^{AB}_{\mu}\sigma_{AB} 
 - i\gamma^{B}e^{\nu}_{B}B_{\mu\nu}]\Psi = 0, \\
 &&\bar{\Psi}\gamma^{C}e^{\mu}_{C}
 [\overleftarrow{\partial}_{\mu} 
 - {i\over4}\omega^{AB}_{\mu}\sigma_{AB} 
 - i\gamma^{B}e^{\nu}_{B}B_{\mu\nu}] = 0.\nonumber
\end{eqnarray}
In the action and Dirac equations above, 
$e^{A}_{\mu}(x)\Bigl(e_{A}^{\mu}(x)\Bigr)$ is the ``vierbein'' 
(and it's inverse) defined by 
$g_{\mu\nu}(x) = \delta_{AB}e^{A}_{\mu}(x)e^{B}_{\nu}(x)$ 
and $e^{A}_{\mu}e_{B}^{\mu} = \delta^{A}_{\ B}$ , 
$e_{A}^{\mu}e^{A}_{\nu} = \delta^{\mu}_{\ \nu}$ 
and $e\equiv(det ~e^{A}_{\mu})$. Thus the Greek indices $\mu,\nu$ refer 
to coordinate basis while the Roman indices A,B = 0,1,2,3 refer 
to non-coordinate basis.  Now $\gamma^{\mu}(x) = e^{\mu}_{A}(x)\gamma^{A}$ 
is the curved spacetime $\gamma$-matrices obeying 
$\{\gamma^{\mu}(x),\gamma^{\nu}(x)\} = -2g_{\mu\nu}(x)$ 
with $\gamma^{A}$ being the usual flat spacetime $\gamma$-matrices.  
Next $(\partial_{\mu} - {i\over4}\omega_{\mu}^{AB}\sigma_{AB})$ is then 
the Lorentz covariant derivative with $\omega_{\mu}^{AB}$ being the spin 
connection and $\sigma_{AB}$ being the $SO(3,1)$ group generator in the 
spinor representation given respectively by

\begin{eqnarray}
 \omega_{\mu B}^{A} &=& -e^{\nu}_{B}
 (\partial_{\mu}e_{\nu}^{A} - \Gamma^{\lambda}_{\mu\nu}e^{A}_{\lambda}),
 \nonumber \\
 \sigma^{AB} &=& {i\over2}[\gamma^{A},\gamma^{B}].
\end{eqnarray}
With this general preparation, now we turn to the examination of 
the existence of fermion zero modes in the background of axionic wormhole 
solutions in Einstein-antisymmetric tensor theory.  As before, we treat 
the problem in two different choices of gauge associated with the time
reparametrization invariance, 
$N(\tau) = 1$ and $N(\tau) = a(\tau)$ one by one. \\
(1) With the gauge choice $N(\tau) = 1$ :
\\
As discussed earlier, in this gauge, the axionic wormhole spacetime 
is described by the Euclidean (spatially-closed) FRW metric  given by

\begin{eqnarray}
 ds^{2} &=& d{\tau}^{2} + a^{2}(\tau)\sigma^{a}{\otimes}\sigma^{a} \\
 &=& g_{\mu\nu}dx^{\mu}dx^{\nu} = \delta_{AB}e^{A}{\otimes}e^{B}\nonumber
\end{eqnarray}
with the scale factor $a(\tau)$ being given by

\begin{eqnarray}
 \tau = {r\over\sqrt{2}}F[cos^{-1}({r\over{a}}),{1\over\sqrt{2}}]
      - \sqrt{2}rE[cos^{-1}({r\over{a}}),{1\over\sqrt{2}}]
      + {1\over{a}}\sqrt{a^{4} - r^{4}}\qquad \Bigl(a(0) = r\Bigr). 
\end{eqnarray}
The non-coordinate basis 1-forms are read off as 

\begin{eqnarray}
 e^{A} = \{e^{0} = d\tau,\  e^{a} = a(\tau)\sigma^{a}\}
\end{eqnarray}
where $\{\sigma^{a}\} ~(a=1,2,3)$ form a basis on the three-sphere $S^{3}$ 
satisfying the $SU(2)$ ``Maurer-Cartan'' structure equation

\begin{eqnarray}
 d\sigma^{a} = {1\over2}\epsilon^{abc}\sigma^{b}{\wedge}\sigma^{c}
\end{eqnarray}
and can be represented in terms of 3-Euler angles $0\le\theta\le\pi$, 
$0\le\phi\le2\pi$ and  $0\le\psi\le4\pi$, parametrizing $S^{3}$

\begin{eqnarray}
 \sigma^{1} &=& cos{\psi}d{\theta} + sin{\psi}sin{\theta}d{\phi},\nonumber\\
 \sigma^{2} &=& sin{\psi}d{\theta} - cos{\psi}sin{\theta}d{\phi}, \\
 \sigma^{3} &=& d{\psi} + cos{\theta}d{\phi}.\nonumber
\end{eqnarray}
Then the associated vierbein and its inverse are found to be 
$($ using $e^{A} = e^{A}_{\mu}dx^{\mu}$ , 
$x^{\mu} = (\tau,\theta,\phi,\psi))$

\begin{eqnarray}
 e^{A}_{\mu} = \left(\matrix
               { 1 & 0          & 0                     & 0 \cr
                 0 & acos{\psi} & asin{\psi}sin{\theta} & 0 \cr
                 0 & asin{\psi} &-acos{\psi}sin{\theta} & 0 \cr
                 0 & 0          & acos{\theta}          & a }
               \right)\quad,\quad
 e_{A}^{\mu} = \left(\matrix
               { 1 & 0          & 0                     & 0 \cr
                 0 & {1\over{a}}cos{\psi} 
                   & {1\over{a}}sin{\psi}
                   & 0 \cr
                 0 & {sin{\psi}\over{asin{\theta}}} 
                   & {-cos{\psi}\over{asin{\theta}}} 
                   & 0 \cr
                 0 & {-sin{\psi}cos{\theta}\over{asin{\theta}}}          
                   & { cos{\psi}cos{\theta}\over{asin{\theta}}}          
                   & {1\over{a}} }
               \right).
\end{eqnarray}
Next, we obtain the spin-connection 1-forms, using the Cartan's 1st structue 
equation (i.e., torsion-free condition)

\begin{eqnarray}
 de^{A} + \omega^{A}_{\ B}{\wedge}e^{B} = 0
\end{eqnarray}
with the help of Maurer-Cartan structure equation given earlier, 
to be (in Euclidean signature)

\begin{eqnarray}
 \omega^{a}_{\mu0} 
 = -\omega^{0}_{\mu{a}} 
 = ({a'\over{a}})e^{a}_{\mu}\quad,\quad
 \omega^{a}_{\mu{b}} 
 = -\omega^{b}_{\mu{a}} 
 = {1\over2a}\epsilon^{abc}e^{c}_{\mu}.
\end{eqnarray}
The KR antisymmetric tensor field $B_{\mu\nu}$ can be given in this 
non-coordinate basis, $B_{AB}$, as well.  Recall that the KR antisymmetric 
tensor field strength giving the axionic wormhole instanton solution 
was given in coordinate basis by 

\begin{eqnarray}
 H_{\mu\nu\lambda} = h(\tau)\epsilon_{\mu\nu\lambda} 
 = {n\over{f_{a}^{2}a^{3}(\tau)}}\epsilon_{\mu\nu\lambda} 
\end{eqnarray}
where $\mu,\nu,\lambda \not=\tau$.  In order to find the associated KR tensor 
field itself, we choose the gauge for which $B_{\mu\nu} = B_{\mu\nu}(\tau)$, 
i.e., $B_{\mu\nu}$ is a function of Euclidean time alone 
and use its relation to its field strength $H = dB$
where
\begin{eqnarray}
 H &=& {1\over3!}h(\tau)\epsilon_{ABC}e^{A}{\wedge}e^{B}{\wedge}e^{C}
    =  {1\over3!}h(\tau)\epsilon_{abc}e^{a}{\wedge}e^{b}{\wedge}e^{c},
    \nonumber\\
 B &=& {1\over2!}B_{AB}(\tau)e^{A}{\wedge}e^{B}
    =  {1\over2!}B_{ab}(\tau)e^{a}{\wedge}e^{b}.
\end{eqnarray}
Note here that we used $e^{A} = e^{A}_{\mu}dx^{\mu}$,
$\epsilon_{ABC} 
= e^{\mu}_{A}e^{\nu}_{B}e^{\lambda}_{C}\epsilon_{\mu\nu\lambda}$ 
and since $\mu,\nu,\lambda\not=\tau$ and the FRW-metric is devoid 
of time-space off-diagonal components, $A,B,C \rightarrow a,b,c \not= 0$. 
Then in non-coordinate basis, the KR antisymmetric tensor field is found to be

\begin{eqnarray}
 B_{ab}(\tau)\epsilon^{acd} 
 = {1\over3}h(\tau)a(\tau)\epsilon^{bcd}
 = {n\over3f^{2}_{a}}{1\over{a^{2}(\tau)}}\epsilon^{bcd}.
\end{eqnarray}
Now, consider the Dirac equation for massless fermion fields 
in the background of KR antisymmetric tensor field and curved spacetime 
obtained earlier

\begin{eqnarray}
 \gamma^{C}e^{\mu}_{C}[\partial_{\mu}
 -{i\over4}\omega^{AB}_{\mu}\sigma_{AB}
 -i\gamma^{B}e^{\nu}_{B}B_{\mu\nu}]\Psi = 0.\nonumber
\end{eqnarray}
For the case at hand,

\begin{eqnarray}
 \gamma^{C}e^{\mu}_{C}\partial_{\mu} &=& \gamma^{0}\partial_{\tau},\nonumber\\
 \gamma^{C}e^{\mu}_{C}\omega^{AB}_{\mu}\sigma_{AB} 
 &=& 6i\gamma^{0}({a'\over{a}}) 
  + i({1\over2a})\epsilon_{abc}\gamma^{a}\gamma^{   b}\gamma^{c}, \\
 \gamma^{\mu}\gamma^{\nu}B_{\mu\nu} &=& \gamma^{a}\gamma^{b}B_{ab}\nonumber
\end{eqnarray}
where we used $\omega^{a0}_{\mu}e^{\mu}_{b} = ({a'\over{a}})\delta^{a}_{b}$,
$\omega^{ab}_{\mu}e^{\mu}_{c} = ({1\over2a})\epsilon^{abc}.$  
Further, assuming that the fermion field depends only on the Euclidean time 
$\tau$ and setting 

\begin{eqnarray}
 \Psi(\tau) = a^{-{3\over2}}(\tau)\tilde{\Psi}(\tau),
\end{eqnarray}
the Dirac equation above reduces to

\begin{eqnarray}
 [\partial_{\tau} + {3\over4}\gamma_{5}{1\over{a(\tau)}} 
 -i\gamma^{0}\gamma^{a}\gamma^{b}B_{ab}(\tau)]\tilde{\Psi}(\tau) = 0
\end{eqnarray}
where $\gamma_{5} = \gamma^{0}\gamma^{1}\gamma^{2}\gamma^{3}$ 
is the Euclidean $\gamma_{5}$-matrix. 
Here, noticing $\gamma^{0}\gamma^{a}\gamma^{b}B_{ab} 
= \gamma_{5}\epsilon^{abc}\gamma^{a}B_{bc}$ and thus if we set, 
using eq.(53),
\begin{eqnarray}
 \epsilon^{abc}\gamma^{a}B_{bc}(\tau) = M{1\over{a^{2}(\tau)}}
\end{eqnarray}
(thus here $M$ is a (4$\times$4) matrix 
whose precise form is not of direct 
relevance for the discussion below), its solution is given by 

\begin{eqnarray}
 \tilde{\Psi}(\tau) 
 = exp[\pm\{{3\over4}\int^{\tau}_{0}{d\tau'\over{a(\tau')}} 
 - iM\int^{\tau}_{0}{d\tau'\over{a^{2}(\tau')}}\}] ~u
\end{eqnarray}
where $u$ denotes the constant basis spinor. Thus the solution to the 
massless Dirac equation is found to be

\begin{eqnarray}
 \Psi(\tau) = {1\over{a^{3\over2}(\tau)}}\tilde{\Psi}(\tau).\nonumber
\end{eqnarray}
Here the $\pm$ signs refer to each of the two chiralities of the fermion 
field. Note that the $\tau$-integration in the exponent is finite 
due to finite integration range and $M$ involves complex matrix. 
Thus owing to the convergence factor $a^{-3/2}(\tau)$ which dies out as 
$\tau\rightarrow\pm\infty$ as we observed earlier, this solution 
to the massless Dirac equation, namely the fermionic zero mode 
is most probably normalizable.  This means the existence of two normalizable 
fermion zero modes. \\
(2) With the gauge choice $N(\tau) = a(\tau)$ : 
\\
In this ``conformal-time gauge'', the Euclidean FRW-metric 
for the $SO(4)$-symmetric axionic wormhole spacetimes 
takes the form given by  

\begin{eqnarray}
 ds^{2} &=& a^{2}(\tau)[d{\tau}^{2} + \sigma^{a}{\otimes}\sigma^{a}] \\
 &=& g_{\mu\nu}dx^{\mu}dx^{\nu} = \delta_{AB}e^{A}{\otimes}e^{B}\nonumber
\end{eqnarray}
with the scale factor $a(\tau)$ being given by

\begin{eqnarray}
 a(\tau) = r[cosh(2\tau)]^{1/2}.
\end{eqnarray}
The non-coordinate basis 1-forms are immediately read off as 

\begin{eqnarray}
 e^{A} = \{e^{0}= a(\tau)d\tau, e^{a}=a(\tau)\sigma^{a}\}
\end{eqnarray}
with $\{\sigma^{a}\} ~(a = 1,2,3)$ again being the left-invariant 1-forms 
on $S^{3}$ satisfying Maurer-Cartan structure equation given earlier.  
Then the associated vierbein and inverse vierbein are found as

\begin{eqnarray}
 e^{A}_{\mu} =a(\tau)\left(\matrix
               { 1 & 0          & 0                     & 0 \cr
                 0 & cos{\psi}  & sin{\psi}sin{\theta}  & 0 \cr
                 0 & sin{\psi}  &-cos{\psi}sin{\theta}  & 0 \cr
                 0 & 0          & cos{\theta}           & 1    }
               \right)\quad,\quad
 e_{A}^{\mu} = {1\over{a(\tau)}}\left(\matrix
               { 1 & 0          & 0                     & 0 \cr
                 0 & cos{\psi}  & sin{\psi}             & 0 \cr
                 0 & {sin{\psi}\over{sin{\theta}}} 
                   & {-cos{\psi}\over{sin{\theta}}} 
                   & 0 \cr
                 0 & {-sin{\psi}cos{\theta}\over{sin{\theta}}}          
                   & { cos{\psi}cos{\theta}\over{sin{\theta}}}          
                   & 1 }
               \right).
\end{eqnarray}
Next, we obtain the spin-connection 1-form using the Cartan's 1st structure 
equation and the $SU(2)$ Maurer-Cartan structure equation given earlier 
and they are

\begin{eqnarray}
 \omega^{a}_{\mu0} 
 = -\omega^{0}_{\mu{a}} 
 = ({a'\over{a^{2}}})e^{a}_{\mu}\quad,\quad
 \omega^{a}_{\mu{b}} 
 = -\omega^{b}_{\mu{a}} 
 = {1\over2a}\epsilon^{abc}e^{c}_{\mu}.
\end{eqnarray}
And as we did before in the case when we chose the gauge $N(\tau) = 1$, 
we can obtain the KR antisymmetric tensor field in non-coordinate basis 
to be

\begin{eqnarray}
 B_{ab}(\tau)\epsilon^{acd} 
 = {n\over3f^{2}_{a}}{1\over{a^{2}(\tau)}}\epsilon^{bcd}
\end{eqnarray}
which turns out to be the same as that in the case with the gauge
choice $N(\tau) = 1$.
Then, again consider the Dirac equation for massless fermion fields 
in the background of axionic wormhole spacetime comprised of the KR 
antisymmetric tensor field and the metric field solution given earlier.  
For the present case in which we choose the gauge $N(\tau) = a(\tau)$, 

\begin{eqnarray}
 \gamma^{C}e^{\mu}_{C}\partial_{\mu} 
 &=& {1\over{a}}\gamma^{0}\partial_{\tau}, \\
 \gamma^{C}e^{\mu}_{C}\omega^{AB}_{\mu}\sigma_{AB} 
 &=& 6i\gamma^{0}({a'\over{a^{2}}}) 
  + i({1\over2a})\epsilon_{abc}\gamma^{a}\gamma^{b}\gamma^{c},\nonumber
\end{eqnarray}
where we used 
$\omega^{a0}_{\mu}e^{\mu}_{b} = ({a'\over{a^{2}}})\delta^{a}_{b}$, 
$\omega^{ab}_{\mu}e^{\mu}_{c} = ({1\over2a})\epsilon^{abc}.$ 
Again, assuming that the fermion field has dependence only on the Euclidean 
time $\tau$ and setting

\begin{eqnarray}
 \Psi(\tau) = a^{-{3\over2}}(\tau)\tilde{\Psi}(\tau),\nonumber
\end{eqnarray}
the Dirac equation becomes 

\begin{eqnarray}
 [\partial_{\tau} + {3\over4}\gamma_{5} 
 -i\gamma^{0}\gamma^{a}\gamma^{b}B_{ab}(\tau)a(\tau)]\tilde{\Psi}(\tau) = 0
 \nonumber
\end{eqnarray}
Now noticing again
$\gamma^{0}\gamma^{a}\gamma^{b}B_{ab} 
= \gamma_{5}\epsilon^{abc}\gamma^{a}B_{bc}$ and using eq.(64), 
if we set 
$\epsilon^{abc}\gamma^{a}B_{bc}(\tau) = M{1\over{a^{2}(\tau)}}$ 
as  before, its solution is given by

\begin{eqnarray}
 \tilde{\Psi}(\tau) 
 = \exp[\pm\{{3\over4}\tau 
 - iM\int^{\tau}_{0}{d\tau'\over{a(\tau')}}\}] u
\end{eqnarray}
with the $\pm$ signs referring to each chirality of the fermion field 
and hence the solution to the massless Dirac equation is found to be

\begin{eqnarray}
 \Psi(\tau) = {1\over{a^{3\over2}(\tau)}}\tilde{\Psi}(\tau).\nonumber
\end{eqnarray}
Now we can invoke the same argument as the one we employed before 
to establish the normalization of this solution to  the massless Dirac 
equation.  Namely, since the $\tau$-integration range in the exponent 
is finite and $M$ involves complex matrix, this fermion zero mode 
is most probably normalizable particularly owing to the obvious convergence 
factor $a^{-3/2}(\tau)$ which dies out as $\tau\rightarrow\pm\infty$. 
And again, this means the existence of two normalizable fermion zero modes.
As commented earlier in this section, the consequence of the existence 
of normalizable fermion zero modes in the background of axionic wormhole 
spacetime may be significant.  Firstly, as has been pointed out by Rey [5]
in the context of wormhole solutions in Einstein-Yang-Mills theory, 
the existence of fermion zero modes would affect wormhole interactions.
Namely, one may expect that, the fermion zero modes, upon integration, 
would yield a long-range confining interaction between the axionic wormholes. 
 Secondly, we have observed in the analysis that the solutions 
to the massless Dirac equation, regardless of the gauge choice 
$N(\tau) =1\ {\rm or}\ a(\tau)$, are symmetric with respect to the chirality 
flip.  And this may signal that the axionic wormhole 
instantons would not induce the chirality-changing  fermion propagation 
in a manner similar to instantons in non-abelian gauge theories typically do. 
As is well-known, in non-abelian gauge theories, chirality-changing fermion 
propagation is attributed to non-trivial instanton configuration or 
non-vanishing instanton 
number which, in turn, is directly related to the chiral anomaly.  
Therefore, the absence of chirality-changing fermion propagation seems 
to imply that the local, axial KR tensor gauge symmetry introduced earlier 
is not anomalous.

\centerline {\rm \bf IV. Quantum Wormholes in Einstein-antisymmetric 
Tensor Theory }

We would like to construct and study a minisuperspace quantum cosmology 
model based on Einstein-KR antisymmetric tensor theory 
(or axionic gravity theory) generally in the presence of the cosmological
constant $\Lambda$ described by the action

\begin{eqnarray}
 S_{EAT} &=& \int_{M}d^{4}x\sqrt{g}
   [{{M_{p}^{2}}\over{16\pi}}R -\Lambda 
   - f_{a}^{2}H_{\mu\nu\lambda}H^{\mu\nu\lambda}]
   + \int_{\partial{M}}d^{3}x\sqrt{h}{{M_{p}^{2}}\over{8\pi}}(K-K_0), \\
 I_{EAT} &=& \int_{M}d^{4}x\sqrt{g}
   [\Lambda - {{M_{p}^{2}}\over{16\pi}}R  
   + f_{a}^{2}H_{\mu\nu\lambda}H^{\mu\nu\lambda}]
   - \int_{\partial{M}}d^{3}x\sqrt{h}{{M_{p}^{2}}\over{8\pi}}(K-K_0)
\end{eqnarray}
in Lorentzian and Euclidean signatures respectively.
As for our approach, we choose to take the avenue of canonical quantum 
cosmology based on Arnowitt-Deser-Misner (ADM)'s (3+1) space-plus-time split 
formulation [13-15].  As usual, then, in order to render the system tractable, 
we reduce the infinite-dimensional superspace down to a 2-dimensional 
minisuperspace by assuming that the 4-dimensional spacetime has the geometry 
of spatially-closed ($k=+1$) FRW-metric.  The geometry of its spatial section 
is, then, that of $S^{3}$ and hence it has $SO(4)$-symmetry.  Then, since 
the spatial geometry is taken to possess the $SO(4)$-symmetry, the matter 
field, i.e., the antisymmetric tensor field (Kalb-Ramond field) defined on it 
should have the same $SO(4)$-symmetry.  Thus we can choose the following 
$SO(4)$-symmetric ans\H atz for the metric and antisymmetric tensor field

\begin{eqnarray}
 ds^{2} &=& \sigma^{2}[-N^{2}(t)dt^{2} + a^{2}(t)d{\Omega}^{2}_{3}] \\
        &=& \sigma^{2}[-N^{2}(t)dt^{2} + a^{2}(t)\sigma^{a}\otimes\sigma^{a}]
         =  \eta_{AB}e^{A}e^{B}\nonumber\\
        {\rm where}\quad e^{A}&=&\{e^{0} = \sigma N(t)dt, ~e^{a} = \sigma a(t)
        \sigma^{a}\}\quad(a = 1,2,3)\nonumber
\end{eqnarray}
and

\begin{eqnarray}
 H &=& {{h(t)}\over{f_{a}}}\epsilon \\
 {\rm where}\quad
 \epsilon &=& {1\over3!}\epsilon_{\mu\nu\lambda}
            dx^{\mu}{\wedge}dx^{\nu}{\wedge}dx^{\lambda} 
          = \sqrt{h}d^{3}x,\nonumber\\
 \qquad\int_{S^{3}} \epsilon &=& \int_{S^{3}}d^{3}x\sqrt{h} 
                      = 2{\pi}^{2}a^{3}(t)\sigma^{4} \nonumber
\end{eqnarray}
where $\sigma^{2}\equiv({2\over3{\pi}M^{2}_{p}})$ has now been introduced 
for convenience and $d{\Omega}^{2}_{3}$ denotes the line element on $S^{3}$.  
Note here that in this quantum treatment, we choose the $SO(4)$-symmetric
ans\H atz for the KR antisymmetric tensor field strength slightly differently
from that in the previous classical treatment in which we took
$H = h(t)\epsilon $.
Again, the left-invariant 1-forms {$\sigma^{a}$} (a = 1,2,3) form a basis 
1-form on $S^{3}$ satisfying the $SU(2)$ Maurer-Cartan structure equation
in eq.(46) and can be represented in terms of 3-Euler angles 
$(\theta, ~\phi, ~\psi)$ parametrizing $S^3$ as given in eq.(47). \\
In order to eventually  write the Lorenzian action in terms of these 
$SO(4)$-symmetric ans\H atz for the metric and antisymmetric tensor field, 
consider

\begin{eqnarray}
 \int{d}^{4}x\sqrt{g} 
 = {\int}dtN(\int_{S^{3}}d^{3}x\sqrt{h}) 
 = 2{\pi}^{2}\sigma^{4}{\int}dtNa^{3}\nonumber
\end{eqnarray} 
and for the choice of the ans\H atz $H_{\mu\nu\lambda} = {h(t)\over{f_{a}}}\epsilon_{\mu\nu\lambda}$

\begin{eqnarray}
 H_{\mu\nu\lambda}H^{\mu\nu\lambda} = \sigma{h^{2}(t)\over{f^{2}_{a}}}.
\end{eqnarray}
Here, first notice that the action term of the Kalb-Ramond antisymmetric 
tensor field $H_{\mu\nu\lambda}H^{\mu\nu\lambda}$ involves no kinetic term 
but only the potential term.
Next, recall that since $H_{\mu\nu\lambda}$ is the field strength for the
Kalb-Ramond antisymmetric tensor $B_{\mu\nu}$, it can be written as
$H = dB$. It, then, immediately follows that $H_{\mu\nu\lambda}$ must satisfy
the Bianchi identity $dH = 0$ which, in our choice of the $SO(4)$-symmetric
FRW-metric, amounts to taking $h(t) = n/a^3(t)$ with $n$ being a constant
to be fixed later.
Namely, recall that the SO(4)-symmetric ans\H atz for the KR antisymmetric
tensor field strength $H_{\mu\nu\lambda}$ (or $h(t)$) given above remains
valid off-shell as well as on-shell as we stressed earlier in the classical 
treatment of the system. And it is because we obtained it simply from
the definition and the Bianchi identity which are certainly bottomline
conditions that should be met in quantum formulations as well. \\
Now, we are ready to write the (Lorentzian) action for Einstein-KR 
antisymmetric tensor 
theory in terms of the $SO(4)$-symmetric ans\H atz for the metric and matter 
field. 

\begin{eqnarray}
 S_{G} &=& {\int}d^{4}x\sqrt{g}[{M^{2}_{p}\over16\pi}R - \Lambda]
       = {1\over2}{\int}dtNa^{3}[-\lambda + \{{1\over{a^{2}}} 
       - ({{\dot a}\over{na}})^{2}\}], \nonumber \\
 S_{AT} &=& {\int}d^{4}x\sqrt{g}[-f^{2}_{a}H_{\mu\nu\lambda}H^{\mu\nu\lambda}]
            \\
        &=& (2{\pi}^{2}\sigma^{4}){\int}dtNa^{3}(-6h^{2}(t))
         = {1\over2}{\int}dtNa^{3}{[-H^{2}(t)]}.  \nonumber
\end{eqnarray} 
Thus

\begin{eqnarray}
 S_{EAT} 
 &=& S_{G} + S_{AT}  \\
 &=& {1\over2}{\int}dtNa^{3}
     [-\lambda + \{{1\over{a^{2}}} - ({{\dot a}\over{na}})^{2}\} - H^{2}(t)]
                   \nonumber
\end{eqnarray}
where we introduced $\lambda \equiv 16\Lambda /9M^4_{p}$ and
$H(t) \equiv \sqrt{24}\pi \sigma^2 h(t) = \sqrt{24}\pi \sigma^2 n/a^3(t)$
and the ``overdot" denotes the derivative with respect to the Lorentzian
time $t$. \\
As mentioned earlier, the definition of $H_{\mu\nu\lambda}$, namely the field 
strength of the KR antisymmetric tensor, $H = dB$, automatically demands it 
to satisfy the Bianchi identity $dH = 0$ which amounts to taking 
$H_{\mu\nu\lambda} = {h(t)\over{f_{a}}}\epsilon_{\mu\nu\lambda} = 
{n\over{f_{a}}}{1\over{a^{3}(t)}}\epsilon_{\mu\nu\lambda}$ in this 
$SO(4)$-symmetric system.  Here, now we fix the constant $n$ such that 
$H(t) = {r^{2}\over{a^{3}(t)}}$, i.e., $r^{2} = \sqrt{24}\pi\sigma^{2}n = 
4\sqrt{6}n/3M^{2}_{p}$.  Then finally the action for the Einstein-KR 
antisymmetric tensor theory takes the form

\begin{eqnarray}
 S_{EAT} 
 &=& {1\over2}{\int}dtNa^{3}
     [-\lambda + \{{1\over{a^{2}}} - ({{\dot a}\over{na}})^{2}\} 
     - {r^{4}\over{a^{6}}}] = {\int}dtL_{ADM}, \nonumber\\
 I_{EAT} 
 &=& {1\over2}{\int}dtNa^{3}
     [\lambda - \{{1\over{a^{2}}} + ({{\dot a}\over{na}})^{2}\} 
     + {r^{4}\over{a^{6}}}] 
\end{eqnarray}
in Lorentzian and in Euclidean signature respectively. \\
Namely, the action for Einstein-KR antisymmetric tensor theory becomes 
effectively that for pure gravity system with an additional potential term 
$\sim{r^{4}/a^{6}}$.  
Next, we obtain the Hamiltonian of this Einstein-KR antisymmetric 
tensor field system via the usual Legendre transformation. To this end,
we first identify the momentum conjugate to the scale factor $a$ as

\begin{eqnarray}
 p_{a} = {{\partial}L_{ADM}\over{\partial}{\dot a}} 
       = {a\over{N}}(-\dot a).
\end{eqnarray}
Thus, from

\begin{eqnarray}
 S_{EAT} 
 &=& {\int}dtL_{ADM} \\
 &=& {\int}dt(p_{a}{\dot a} - H_{ADM})
 = {\int}[p_{a}da - (NH_{0} + N_{i}H^{i})dt]\nonumber
\end{eqnarray}
where $H_{ADM} = NH_{0} + N_{i}H^{i}$,
it follows that
\begin{eqnarray}
 {{\delta}S_{EAT}\over{\delta}N}
 &=& {1\over2}a^{3}[-\lambda - {r^{4}\over{a^{6}}} + {1\over{a^{2}}} 
     + {{\dot a}^{2}\over{N^{2}a^{2}}}] \\
 &=& {1\over{2a}}[p^{2}_{a} - \{{\lambda}a^{4} - a^{2} + 
 {r^{4}\over{a^{2}}}\}].     \nonumber
\end{eqnarray}
Namely, the Hamiltonian of the system is found to be    
\begin{eqnarray}
 H_{0} = -{{\delta}S_{EAT}\over{\delta}N}
 &=& {1\over{2a}}[-p^{2}_{a} 
     + \{{\lambda}a^{4} - a^{2} + {r^{4}\over{a^{2}}}\}]\nonumber\\
 &\equiv& {1\over{2a}}[-p^{2}_{a} + U(a)] \\
 {\rm where}\quad U(a) &=& {\lambda}a^{4} - a^{2} + {r^{4}\over{a^{2}}}.
 \nonumber
\end{eqnarray}
General relativity is one of the most well-known constrained system.  
The invariance of the system under the 4-dim. diffeomorphisms (consisting of 
the time-reparametrization and the 3-dim. general coordinate transformations 
of the spacelike hypersurface) leads to the emergence of 4-constraint 
equations.  Of them, we need not explicitly impose the 3-momentum constraint 
equations since we already have taken the $N^{i} = 0$ gauge which amounts to 
assuming the $SO(4)$-symmetric spatially-closed FRW-metric.  
Thus, we only need to impose the 
Hamitonian constraint $H_{0} = 0$.  The classical Hamiltonian constraint now 
reads

\begin{eqnarray}
 H_{0} 
 = {1\over{2a}}[-p^{2}_{a} 
 + \{{\lambda}a^{4} - a^{2} + {r^{4}\over{a^{2}}}\}] = 0.
\end{eqnarray}
Now, in order to quantize this Einstein-KR antisymmetric tensor system, we 
need to turn to the ``Dirac quantization procedure'' for the constrained 
system.  According to the Dirac quantization procedure, the invariance in the 
action of the theory under the 4-dim. diffeomorphism is secured by demanding 
that the physical (universe) wave function $\Psi $ be annihilated by 
``operator versions'' of the 4-constraints.  Therefore, the classical 
Hamiltonian constraint above turns into its quantum version, namely the 
Wheeler-DeWitt equation given by

\begin{eqnarray}
 {\hat H_{0}}(p_{a} = -i{{\partial}\over{\partial}a})\Psi[a] = 0.
\end{eqnarray}
In order to obtain the correct form of this Wheeler-DeWitt equation, we first 
examine the structure of the classical Hamiltonian

\begin{eqnarray}
 H_{0} 
 &=& {1\over{2a}}[-p^{2}_{a} 
 + \{{\lambda}a^{4} - a^{2} + {r^{4}\over{a^{2}}}\}] = 0.\nonumber\\
 &=& T + V \equiv {1\over2}G^{\alpha\beta}\Pi_{\alpha}\Pi_{\beta} + V.
\end{eqnarray}
Here, one can readily read off the ``minisuperspace metric'' $G_{\alpha\beta}$
as 

\begin{eqnarray}
 G_{\alpha\beta} = -a\delta_{\alpha\beta}\quad,\quad
 G^{\alpha\beta} = -{1\over{a}}\delta^{\alpha\beta}
\end{eqnarray}
with $\gamma^{\alpha} = a$ and $\Pi_{\alpha} = p_{a}$ being the 
minisuperspace variable and its conjugate momentum respectively.  
Now, by the usual substitution, 

\begin{eqnarray}
 G^{\alpha\beta}\Pi_{\alpha}\Pi_{\beta}\quad\rightarrow\quad -\nabla^{2}
\end{eqnarray}
with $ \nabla^{2} = {1\over\sqrt{G}}{{\partial}\over{\partial}\gamma^{\alpha}}
(\sqrt{G}G^{\alpha\beta}{{\partial}\over{\partial}\gamma^{\beta}}) 
= -{1\over{a}}{{\partial}\over{\partial}a}({{\partial}\over{\partial}a})$, 
finally we arrive at the Wheeler-DeWitt equation

\begin{eqnarray}
 {\hat H_{0}}\Psi = 
 {1\over2}\Bigl[{1\over{a}}{{\partial}\over{\partial}a}
 ({{\partial}\over{\partial}a}) + {1\over{a}}U(a)\Bigr]\Psi[a] = 0.
\end{eqnarray}
Note, here, that the minisuperspace metric $G_{\alpha\beta}(\gamma)$ is 
generally a function of minisuperspace variables $\gamma^{\alpha}$.  
Therefore, in passing from classical to quantum version there arises the 
``ambiguity in operator ordering'' problem.  Thus, although it is not the 
most general form, by rewriting 

\begin{eqnarray}
 {{\partial}\over{\partial}a}({{\partial}\over{\partial}a})
 \quad\rightarrow\quad
 {1\over{a^{p}}}{{\partial}\over{\partial}a}(a^{p}{{\partial}\over{\partial}a})
\end{eqnarray}
as suggested by Hartle and Hawking [14], one can partly encompass 
the ``operator-ordering''problem.  Finally, the Wheeler-DeWitt equation 
generally takes the form

\begin{eqnarray}
 {1\over2}\Bigl[{1\over{a^{p}}}{{\partial}\over{\partial}a}
 (a^{p}{{\partial}\over{\partial}a}) + U(a)\Bigr]\Psi[a] = 0
\end{eqnarray}
where ``$p$'' denotes an index representing the ambiguity in 
``operator-ordering'' and $U(a) = (\lambda a^{4} - a^{2} + 
{r^{4}\over{a^{2}}})$.  First of all, in order to have some insight 
into the behavior of the solution of this Wheeler-DeWitt equation we 
assign the ``normal'' sign to the ``kinetic'' energy term to get

\begin{eqnarray}
 {1\over2}\Bigl[{-1\over{a^{p}}}{{\partial}\over{\partial}a}
 (a^{p}{{\partial}\over{\partial}a}) + \tilde{U}(a)\Bigr]\Psi[a] = 0.
\end{eqnarray}
Then the ``potential'' energy can be identified with

\begin{eqnarray}
 \tilde{U}(a) = -U(a) = (a^{2} - {\lambda}a^{4} - {r^{4}\over{a^{2}}}).
\end{eqnarray}
The Fig.3(4) given displays the plot of ``potential''energy as it appears 
in the WD equation in the presence (absence) of the cosmological constant 
$\lambda = 16\Lambda/9M^{2}_{p}$.  Both figures show that due to the 
contribution to the potential energy, $(-{r^{4}/a^{2}})$, coming from 
the KR antisymmetric tensor sector of the theory, the potential develops 
an ``abyss'' in the small-$a$ region regardless of the presence or absence 
of the cosmological constant term.  Since the WD equation implies that the 
total energy of the gravity-matter system is zero, $E = 0$, the emergence 
of the abyss in the small-$a$ region of the potential readily reveals the fact 
that the universe wave function $\Psi[a]$ should be a highly oscillating 
function of $a$ there.  And this small-$a$ behavior of the universe wave 
function,
 namely the enormous oscillation for small scale factor $a$ appears to signal 
the existence of ``quantum wormhole'' as well as other types of spacetime 
fluctuations in the small-$a$ region of the superspace and hence seems 
consistent with the existence of classical wormhole solution in this 
Einstein-KR antisymmetric tensor theory as we have seen in the earlier 
sections.   Obviously, the most straightforward way of confirming the 
possible existence of ``quantum wormholes'' is to solve the WD equation given 
above for the universe wave functiion.  Unfortunately, exact, analytic 
solutions to the WD equation in the presence of the cosmological constant 
are not available (exact solutions to the WD equation even for de Sitter 
spacetime pure gravity are not available either [14,15]).  In the absence of the 
cosmological constant, $\lambda = 0$, however, an exact, analytic solution 
to the WD equation is available.  There, of course, is a well-known issue 
of initial or boundary condition for the universe wave functioin.  
The WD equation, which plays the role of Schr\H odinger-type equation for 
the universe state, is a second order hyperbolic functional differential 
equation describing the evolution of the universe wave functioin in 
superspace.  Thus the WD equation, in general, has a large number of 
solutions and in order to have any predictive power, one needs initial 
or boundary conditions to pick out just one solution by, for instance, 
giving the value of universe wave function at the boundary of the superspace 
on which it is defined.  Thus far, a number of different proposals for the 
law of initial or boundary conditioins have been put forward.  
And among them, ``no-boundary proposal'' of Hartle and Hawking (HH) [14] and 
``tunnelling boundary condition'' due to Vilenkin [15] are the ones 
which are the most comprehensive and the most extensively studied.  
If stated briefly, the no-boundary 
proposal by HH [14] is based on the philosophy that the quantum state of the 
universe is the closed cosmology's version of ``ground state'' or ``state 
of minimal excitation'' and the wave function of this ground state is given 
by an Euclidean sum-over-histories.  Next, Vilenkin's tunnelling boundary 
condition [15] can be best stated in ``outgoing modes'' formulation 
which governs 
the behavior of the solutions to the WD equation at boundaries of the 
superspace.  Namely, according to this proposal, at ``singular boundaries''
(such as the region of zero 3-metric and infinite 3-curvature 
($\sqrt{h}\rightarrow 0$) of superspace), the universe wave function should 
consist solely of outgoing modes carrying flux out of superspace.  
In practice, these two proposals for the law of initial or boundary 
conditions essentially aim at giving particular boundary conditions on the 
universe wave function at singular boundaries ($\sqrt{h}\rightarrow 0$) 
of the superspace which, presumably, are the points where the universe 
(or the universe wave function) has started.  These boundary conditions, 
then, determine the behaviors of the universe wave function like how the 
universe nucleated (from ``nothing'') and then following which line it has 
subsequently evolved to the present one.  Indeed, in the minisuperspace model 
(where the ``singular'' boundary corresponds to the point $a\rightarrow 0$) 
and within the context of the semiclassical approximation, these two 
proposals have been successfully applied to and tested for simple systems 
such as de Sitter spacetime pure gravity or a scalar field theory coupled 
minimally to gravity concretely demonstrating the ways how the universe 
nucleates and then subsequently evolves.  Therefore, in view of this, 
applying these boundary conditions on the universe wave function 
(particular Vilenkin's tunnelling boundary condition) to the present case is, 
in many respects, irrelevant or, at least, awkward since we are supposed to 
determine the universe wave function in the small-$a$ region, namely on the 
boundary itself of the (mini)superspace.  Namely, for the case at hand, 
i.e., in the Einstein-KR antisymmetric tensor theory in the absence of the 
cosmological constant, the shape of the potential (given in Fig.4) as it 
appears in the WD equation ``traps'' the universe (namely the value of the 
scale factor, $a$) within a small-$a$ region and we would like to determine 
the state of the universe in this region.  Consequently, questions like how 
the universe nucleates or how it subsequently evolves are irrelevant.  
(Of course, HH's no-boundary proposal might still be relevant to be 
considered even for the present case since its formulation is based on the 
philosophy with wide applicability to general situations.) 
Therefore in the following, we present an exact, analytic 
solution to the WD equation in the absence of the cosmological constant that 
is obtained by directly integrating the WD equation and is not constructed 
from any of these boundary conditions. Further, since this exact solution is 
a mathematical one, later we shall impose some conditions on the parameters 
involved in the solution in order for the resulting universe wave function 
to have physically relevant interpretations.  Now consider the WD equation 
as given earlier but in this time in the absence of the cosmological 
constant.  And in the following discussion, we shall set the constant $n$ 
appearing in the $SO(4)$-symmetric ans\H atz for the KR field strength 
$h(t) = n/a^{3}(t)$ to be $n=(\sqrt{24}\pi\sigma^{2})^{-1}$ so that the 
parameter $r^{2} = \sqrt{24}\pi\sigma^{2}n$ becomes unity.  In fact, 
this rescaling amounts to taking natural unit in which $M^{2}_{p} = 1$. 
The WD equation, then, takes the form
\begin{eqnarray}
 \Bigl[{{\partial^{2}}\over{\partial}a^{2}} 
 + {p\over{a}}{{\partial}\over{\partial}a} 
 - a^{2} + {1\over{a^{2}}}\Bigr]\Psi[a] = 0.
\end{eqnarray}  
An exact solution to this ordinary, second-order differential equation 
is given by

\begin{eqnarray}
 \Psi[a] 
 &=& Ca^{({{1-p}\over2})}Z_{\nu}({i\over2}a^{2})\nonumber\\
 &=& a^{({{1-p}\over2})}\Bigl[AJ_{\nu}({i\over2}a^{2}) 
     + BN_{\nu}({i\over2}a^{2})\Bigr]\nonumber
\end{eqnarray}
where $Z_{\nu}(z)$ is the Bessel function satisfying the Bessel equation 
and hence is generally given by the linear combination of the Bessel function 
of the 1st kind $J_{\nu}(z)$ which is regular for $z\rightarrow 0$ and the 
Bessel function of the 2nd kind (i.e., Neumann function) $N_{\nu}(z)$ which 
is regular for $z\rightarrow \infty$.  And here the order of the Bessel 
function is given by $\nu = {1\over4}\sqrt{(p+1)(p-3)}$ with $p$ being the 
suffix indicating the ambiguity in ``operator-ordering''.  $A, ~B$ and $C$ are 
arbitrary constant coefficients yet.  Note that the structure of the 
WD equation 
above indicates that we are dealing with an one-dimensional 
Schr\H odinger-type equation with the total energy $E=0$ and the potential 
given in Fig.4.  Thus we expect that the physical solution, 
i.e., the universe wave function $\Psi[a]$ possesses a highly oscillating 
behavior for $a\rightarrow 0$ whereas a rapidly damping behavior for large 
$a$.  The exact solution to the WD equation given above is yet just 
a mathematical solution.  Now, we would like to turn it into a physical 
universe wave function by imposing physical conditions, namely by demanding 
that it satisfy appropriate asymptotic behaviors stated above.  Fortunately, 
the exact solution above involves an undetermined parameter $p$ which is the 
index representing the operator-ordering ambiguity.  Since this parameter 
$p$ controls the behavior of the solution, namely the Bessel function, 
we shall be able to obtain a physical solution by fixing its value in such 
a way that with certain values of $p$ the exact solution takes on expected 
asymptotic behavior stated above.  Therefore, to this end, we carefully 
consider the asymptotic behaviors of the Bessel function.  First for  
$z\rightarrow 0$, $Z_{\nu}(z)= J_{\nu}(z)\sim {z^{\nu}}/{2^{\nu}{\nu}!}$.
  Thus for  $a\rightarrow 0$ 

\begin{eqnarray}
 Z_{\nu}({i\over2}a^{2})\sim {1\over{2^{\nu}\nu!}}({i\over2}a^{2})^{\nu}
                        \sim a^{2\nu}.
\end{eqnarray}
Now, in order for the universe wave function 
$\Psi[a]{\sim}a^{({{1-p}\over2})}a^{2\nu}$ 
with $\nu = {1\over4}\sqrt{(p+1)(p-3)}$ in the region of small-$a$ to have 
enormously oscillating behavior, the order $\nu$ should be imaginary, 
$\nu = i|\nu|$ which amounts to choosing $-1<p<3$.  
Consequently, the behavior of the 
universe wave function for  $a\rightarrow 0$ is given by 

\begin{eqnarray}
 \Psi[a]{\sim}a^{({{1-p}\over2})}\exp{[i2|\nu|{\rm ln}a]}
\end{eqnarray}
where $|\nu| = {1\over4}\sqrt{|(p+1)(p-3)|}$ with  $-1<p<3$. Apparently, 
this universe wave function possesses highly oscillatory behavior for  
$a\rightarrow 0$ and hence possibly represents a wave function of small scale 
spacetime fluctuations including wormholes.  Note also that $\Psi[a] = 0$ 
at $a=0$, namely it becomes regular for $-1<p<1$. Next, for $z\rightarrow$ 
large, 
$J_{\nu}(z){\sim}\sqrt{{2\over{\pi}z}}cos(z - {\nu\pi\over2} - {\pi\over4})$ 
and 
$N_{\nu}(z){\sim}\sqrt{{2\over{\pi}z}}sin(z - {\nu\pi\over2} - {\pi\over4})$, 
thus $Z_{\nu}(z){\sim}{1\over\sqrt{z}}e^{{\pm}iz}.$  Therefore, 
for $a\rightarrow$ large

\begin{eqnarray}
 Z_{\nu}({i\over2}a^{2}){\sim}{1\over{a}}e^{{\pm}{1\over2}a^{2}}
\end{eqnarray} 
and hence the universe wave function behaves in the region of large-$a$ as

\begin{eqnarray}
 \Psi[a]{\sim}a^{-({{1+p}\over2})}e^{-{1\over2}a^{2}}
\end{eqnarray}
where we choose the minus sign in the exponent since it is the physically 
relevant one. Namely, the universe wave function possesses rapidly damping 
behavior for $a\rightarrow$ large and this is exactly what we expected.  
Finally the physically relevant solution to the WD equation in the 
Einstein-KR antisymmetric tensor theory (in the absence of the cosmological 
constant), namely the universe wave function possibly of the quantum 
wormholes is given by 

\begin{eqnarray}
 \Psi[a] = C a^{({{1-p}\over2})}Z_{i|\nu|}({i\over2}a^{2})
         = C a^{({{1-p}\over2})}J_{i|\nu|}({i\over2}a^{2})
\end{eqnarray}
where  $|\nu| = {1\over4}\sqrt{|(p+1)(p-3)|}$ with  $-1<p<1$ and we dropped 
the Neumann function term demanding that the universe wave function remain 
finite for $a\rightarrow 0$.  We believe that 
the solution to the WD equation given in eq.(94) would represent quantum 
wormhole 
spacetimes.  Now, we would like to fortify this belief of ours in 
an unambiguous manner.  As had been advocated by Hawking and Page [12], 
in order 
for a solution to a WD equation to represent quantum wormholes, it should 
obey certain boundary conditions.  And the appropriate boundary conditions 
for wormhole wave functions seem to be that they are damped, say, 
exponentially for large 3-geometries ($\sqrt{h}\rightarrow\infty$) and are 
regular in some suitable way when the 3-geometry collapses to zero 
($\sqrt{h}\rightarrow 0$).  Particularly in the context of FRW minisuperspace 
model, large 3-geometries correspond to  $a\rightarrow$ large limit and the 
3-geometry collapsing to zero corresponds to $a\rightarrow 0$ limit.  And the 
damping behavior of the universe wave function at large -$a$ indicates that 
there are no gravitational excitations asymptotically and hence it represents 
asymptotically Euclidean spacetime while its regularity at $a=0$ indicates 
that it is nonsingular.  Therefore the solution to a WD equation obeying 
these boundary conditions must correspond to wormholes that connect two 
asymptotically Euclidean regions.  Now we turn to our solution to the WD 
equation in Einstein-KR antisymmetric tensor theory given in eq.(94)
and see if it indeed obeys these boundary conditions.  Although we took the
 natural unit in which $M^{2}_{p} = 1$ and rescaled such that 
$r^{2} = \sqrt{24}\pi\sigma^{2}n = 4\sqrt{6}n/3M^{2}_{p}$ takes the value of 
unity, we recover this length parameter for the moment.  
Firstly for $a\rightarrow 0$ or more concretely for $0<a<r$, the universe 
wave function behaves like 
$\Psi{\sim}a^{({{1-p}\over2})}\exp{[i2|\nu|{\rm ln}a]}$ thus it oscillates 
infinitely and hence would correspond to initial or final spacetime 
singularities.  In addition for $-1<p<1$, this solution is regular, 
i.e., $\Psi[a] = 0$ at $a=0$.  Secondly for $a\rightarrow$ large or more 
concretely for $a>r$, the universe wave function behaves as  
$\Psi{\sim}a^{-({{1+p}\over2})}e^{-a^{2}/2}$ thus it damps rapidly 
enough and thus represents asymptotically Euclidean regions.  
Namely our solution does satisfy the boundary conditions for wormhole wave 
functions.  Besides, the lower bound $a=r$ of the oscillating solution on the 
radius $a$ of $S^{3}$ and the existence of the conserved axion current flux 
through the $S^{3}$ of the solution, i.e.,

\begin{eqnarray}
 \int_{S^{3}}H =\int_{S^{3}}{h(t)\over{f_{a}}}\epsilon
               = {2\pi^{2}n\over{f_{a}}}
\end{eqnarray}
indicates that indeed our solutioin describes a wormhole connecting two 
asymptotically Euclidean regions.  Finally, since our solution oscillates 
infinitely near $a=0$, it would be expressible as an infinite sum of 
a discrete family of solutions to the WD equation that are well-behaved both 
at zero radius (i.e., ``regularity'') and at infinity (i.e., ``damping'').  
And this completes the study of the solution to the WD equation in 
Einstein-KR antisymmetric tensor theory in the absence of the cosmological 
constant.  It is interesting to note that our knowledge on the nature of the 
universe wave function in the absence of the cosmological constant developed 
thus far may, in turn, enable us to construct the solution to the WD equation 
in the presence of the cosmological constant at least approximately yet quite 
systematically.  Thus in what follows, we shall turn to this problem.  
Now, we go back and consider the WD equation in Einstein-KR antisymmeric 
tensor theory in the presence of the cosmological constant

\begin{eqnarray}
 {1\over2}\Bigl[-{1\over{a^{p}}}{{\partial}\over{\partial}a}
 (a^{p}{{\partial}\over{\partial}a}) 
 + (a^{2} - {\lambda}a^{4} - {1\over{a^{2}}})\Bigr]\Psi[a] = 0.
\end{eqnarray}     
With the potential energy, $\tilde{U}(a) = (a^{2} - {1\over{a^{2}}})$, 
as it appears in the WD equation in the absence of the cosmological constant, 
we now know that the solution to the Wheeler-DeWitt (WD) equation represents 
a quantum wormhole and particularly near $a=0$, it oscillates infinitely and 
hence corresponds to large spacetime fluctuations.  Therefore this 
observation plus the shape of the full potential energy, $\tilde{U}(a) = 
(a^{2} -{\lambda}a^{4} - {1\over{a^{2}}})$ in the presence of the 
cosmological constant as was depicted in Fig.3 suggest that the solution 
to the WD equation above would describe the state of the universe that 
undergoes ``large spacetime fluctuatioins for very small-$a$''$\rightarrow$ 
``spontaneous nucleation (quantum tunnelling) of the universe from nothing 
in a de Sitter geometry'' $\rightarrow$ ``subsequent, mainly classical 
evolution of the universe for large-$a$''.  Namely, if we are willing to 
accept (${\partial}/{\partial}a$) as the timelike killing field in the 
(mini)superspace, the Einstein-antisymmetric tensor theory in the presence 
of the cosmological constant appears to serve as a simple yet interesting 
model which provides a comprehensive overall picture of entire universe's 
history from the deep quantum domain all the way to the essentially classical 
domain.  Then coming back to a practical problem, now we wish to construct 
the approximate solutions to the WD equation in eq.(96).  Clearly, the 
behavior of the 
solution for very small-$a$ will be determined by the wormhole wave function 
obtained in the present work while the behavior for intermediate-to-large-$a$ 
regions will be governed by the de Sitter space universe wave function.  
And as mentioned earlier, the de Sitter space universe wave functions has 
been constructed and extensively studied in the context of semiclassical 
approximation with the choice of both HH's no-boundary proposal [14] and 
Vilenkin's tunnelling boundary condition [15].  Thus to work out this 
idea, we 
consider the WD equation in two regions of interest in the minisuperspace.  
Firstly for very small-$a$ region, the WD equation reduces to eq.(89)
and the exact solution to this equation representing particularly the large spacetime fluctuations near $a=0$ is given by

\begin{eqnarray}
 \Psi_{I}[a] &=& a^{({{1-p}\over2})}J_{i|\nu|}({i\over2}a^{2}) \\
         &\rightarrow& a^{({{1-p}\over2})}\exp{[i2|\nu|{\rm ln}a]}\qquad 
         ({\rm for}\quad a\rightarrow 0)\nonumber\\
         &\rightarrow& a^{-({{1+p}\over2})}e^{{\pm}{1\over2}a^{2}}\qquad 
         \qquad({\rm for}\quad a\rightarrow {\rm large})\nonumber
\end{eqnarray}
where $|\nu| = {1\over4}\sqrt{|(p+1)(p-3)|}$ with $-1<p<1$.  Secondly for 
intermediate-to-large-$a$ regions, the WD equation reduces to

\begin{eqnarray}
 \Bigl[{{\partial^{2}}\over{\partial}a^{2}} 
 + {p\over{a}}{{\partial}\over{\partial}a} 
 - a^{2} + {\lambda}a^{4}\Bigr]\Psi[a] = 0.\nonumber  
\end{eqnarray}
As mentioned, the semiclassical approximation to the solutions of this de 
Sitter space WD equation has been thoroughly studied.  First HH's no-boundary 
wave function [14] is given by
 
\begin{eqnarray}
 \Psi^{HH}_{II}[a]
 &=& a^{-({{1+p}\over2})}\exp{\Bigl[{1\over3\lambda}
     \{1-(1-{\lambda}a^{2})^{3/2}\}\Bigr]}\qquad
     ({\rm for}\quad{\lambda}a^{2}<1)\nonumber\\
   &&\rightarrow a^{-({{1+p}\over2})}e^{{1\over2}a^{2}}
     \quad({\lambda}a^{2}<<1), \\
 &=& a^{-({{1+p}\over2})}\exp{[{1\over3\lambda}]}
     2cos\Bigl[{({\lambda}a^{2} - 1)^{3/2}\over3\lambda} - {\pi\over4}\Bigr]
     \qquad({\rm for}\quad{\lambda}a^{2}>1)\nonumber\\
   &&\rightarrow a^{-({{1+p}\over2})}
     \Bigl[e^{i{\sqrt{\lambda}\over3}a^{3}} 
     + e^{-i{\sqrt{\lambda}\over3}a^{3}}\Bigr]
     \quad({\lambda}a^{2}>>1).\nonumber
\end{eqnarray}
This HH's no-boundary wave function consists of both ``ingoing''(contracting) 
and ``outgoing''(reexpanding) modes in the classical-allowed region 
(${\lambda}a^{2}>1$) which, then, decreases exponentially as it moves towards 
smaller values of $a$ in the classically-forbidden region (${\lambda}a^{2}<1$).
Next, Vilenkin's tunnelling wave function [15] is given by

\begin{eqnarray}
 \Psi^{T}_{II}[a]
 &=& a^{-({{1+p}\over2})}(1-{\lambda}a^{2})^{-1/4}
     \exp{\Bigl[-{1\over3\lambda}\{1-(1-{\lambda}a^{2})^{3/2}\}\Bigr]}\qquad
     ({\rm for}\quad{\lambda}a^{2}<1)\nonumber\\
   &&\rightarrow a^{-({{1+p}\over2})}e^{-{1\over2}a^{2}}
     \quad({\lambda}a^{2}<<1), \\
 &=& a^{-({{1+p}\over2})}e^{i{\pi\over4}}({\lambda}a^{2}-1)^{-1/4}
     \exp{\Bigl[-{1\over3\lambda}\{1+i({\lambda}a^{2}-1)^{3/2}\}\Bigr]}\qquad
     \qquad({\rm for}\quad{\lambda}a^{2}>1)\nonumber
\end{eqnarray}
This Vilenkin's tunnelling wave function exponentially decreases as it moves 
from small toward larger values of $a$ (i.e., emerges out of the potential 
barrier via ``quantum tunnelling'') and then upon escaping the barrier, 
it consists solely of ``outgoing'' (expanding) mode in the classically-allowed 
region (${\lambda}a^{2}>1$). Now we are ready to write down approximate 
solutions to the WD equation in the presence of the cosmological constant by 
putting these pieces altogether.  To do so let us denote the smaller and 
larger roots of the equation

\begin{eqnarray}
 \tilde{U}(a) = a^{2} - {\lambda}a^{4} - {1\over{a^{2}}} = 0\nonumber
\end{eqnarray}
by $r_{-}$ and $r{+}$ respectively.  Note that this equation has positive 
roots provided $\Lambda<({3\over8})^{2}M^{4}_{p}$ or $\lambda<{1/4}$ and the 
two roots are given by $r^{4}_{\pm} = {1\over2\lambda}
[1\pm\sqrt{1-4\lambda}]$. \\        
(1) With the choice of HH's no-boundary wave function :

\begin{eqnarray}
 \Psi[a]
 &=& \Psi_{I}[a]\qquad\qquad({\rm region}\ I: 0<a<r_{+}),\nonumber\\
 &=& \Psi^{HH}_{II}[a]\qquad\qquad({\rm region}\ II: r_{-}<a<{\infty}).
 \nonumber
\end{eqnarray}
(2) With the choice of Vilenkin's tunnelling  wave function :

\begin{eqnarray}
 \Psi[a]
 &=& \Psi_{I}[a]\qquad\qquad({\rm region}\ I: 0<a<r_{+}),\nonumber\\
 &=& \Psi^{T}_{II}[a]\qquad\qquad({\rm region}\ II: r_{-}<a<{\infty}).
 \nonumber
\end{eqnarray}
The two kinds of universe wave functions corresponding to the two 
different choices of the boundary conditions are plotted in Fig 5. and 6 
respectively.

\centerline {\rm \bf V. Discussions}

Now we summarize the motivation and the results of the present work. \\
We revisited, in this work, the Einstein-KR antisymmetric tensor theory
considered first by Giddings and Strominger [4] which is a classic system
known to admit classical, Euclidean wormhole instanton solution. 
Although the classical wormhole instanton as a solution to the classical
field equations and much of its effects on low energy physics have been
studied extensively in the literature, some of important aspects of the 
classical wormhole physics such as the existence and the physical
implications of fermion zero modes in the background of classical
axionic wormhole spacetime has not been addressed. Moreover, since this
Einstein-KR antisymmetric tensor system admits classical wormhole
solutions, one may wonder if there is any systematic way of exploring
the existence and the physics of ``quantum'' wormholes in the same
theory. The present work attempted to deal with these kinds of yet
unquestioned issues. And firstly, in order to investigate the existence
of the fermion zero modes and their physical implications, we followed
the formulation taken by Hosoya and Ogura [5] in their study of classical
wormhole instantons in Einstein-Yang-Mills theory. And to do so, we 
needed to introduce the fermion-KR antisymmetric tensor field
interactions possessing, of course, the general covariance and the
local gauge-invariance. And the result was that regardless of the 
gauge choices associated with the time reparametrization invariance,
i.e., $N(\tau) = 1$ or $a(\tau)$, there are two normalizable
fermion zero modes. As we mentioned in the text, the existence of
fermion zero modes would affect wormhole interactions. Namely, the
fermion zero modes, upon integration, would yield a long-range
confining interaction between the axionic wormholes. Next, the
fermion zero modes, i.e., the solutions to the massless Dirac
equation, are symmetric with respect to the chirality flip. And this
may signal that the axionic wormhole instantons would not induce the
chirality-changing fermion propagation unlike the typical instantons
in non-abelian gauge theories. Secondly, in order to explore the
quantum wormholes in this system systematically, we worked in the
context of canonical quantum cosmology and followed Hawking and Page [12]
to define the quantum wormhole as a state or an excitation represented
by a solution to the Wheeler-DeWitt equation satisfying a certain 
wormhole boundary condition. Particularly, in the minisuperspace quantum
cosmology model possessing  SO(4)-symmetry, an exact, analytic solution
to the Wheeler-DeWitt equation satisfying the appropriate wormhole
boundary condition was found in the absence of the cosmological
constant. Thus we confirmed our expectation that the Einstein-KR
antisymmetric system admits quantum wormholes as well as classical
wormholes. Further, we pointed out that the minisuperspace quantum
cosmology model based on this Einstein-KR antisymmetric tensor theory
in the presence of the cosmological constant may serve as an simple
yet interesting system displaying an overall picture of entire
universe's history from the deep quantum domain all the way to the
classical domain. \\
As we have stressed in the text, the essential point that allowed us to
explore, in a concrete manner, the quantum wormhole in the context of the
minisuperspace quantum cosmology model was the following observation.
In their original work, Giddings and Strominger [4] looked for an economical
way of solving the coupled Einstein-KR antisymmetric tensor field 
equations. They found out that the classical Euler-Lagrange's equation
of motion $d^{\ast}H = 0$ and the Bianchi identity $dH = 0$ can be
simultaneously satisfied if one takes the SO(4)-symmetric ans\H atz for
the KR antisymmetric tensor field strength as $H_{\mu\nu\lambda} =
{n\over f^2_{a}a^3}\epsilon_{\mu\nu\lambda}$ which, in turn, reduces the
Einstein equation to that of the scale factor $a(\tau)$ alone.
However, we realized in this work that even without imposing the
on-shell condition (i.e., the classical field equation), one can ``derive''
$H_{\mu\nu\lambda} = {const.\over a^3}\epsilon_{\mu\nu\lambda}$ just
from the definition $H = dB$ and the Bianchi identity $dH = 0$.
Therefore this SO(4)-symmetric ans\H atz for the KR antisymmetric tensor
field strength $H_{\mu\nu\lambda}$ remains valid even off-shell as well as
on-shell and hence can be used in the quantum treatment of the 
Einstein-KR antisymmetric tensor field system. Consequently, the 
Wheeler-DeWitt equation in the context of the canonical quantum cosmology
becomes a Schr\H odinger-type equation of the minisuperspace variable
$a$ (the scale factor) alone and can be solved exactly particularly in
the absence of the cosmological constant. \\
Finally, our study of quantum wormholes in this work appears to demonstrate
that, after all, the Einstein-KR antisymmetric tensor theory is a simple
(although it is a truncated system of a more involved, fundamental
string theory) but fruitful system which serve as an arena in which we can 
envisage quite a few exciting aspects of quantum gravitational phenomena.

\centerline {\rm \bf Acknowledgements}
This work was supported in part by Korea Research Foundation and Basic
Science Research Institute (BSRI-97-2427) at Ewha Women's Univ.

\newpage
\vspace*{2cm}

\centerline {\rm \bf References}

\begin{description}

\item {[1]} E. Baum, Phys. Lett. {\bf B133}, 185 (1983) ; 
            S. W. Hawking, Phys. Lett. {\bf B134}, 403 (1984).
\item {[2]} S. Giddings and A. Strominger, Nucl. Phys. {\bf B307},
            854 (1988).
\item {[3]} S. Coleman, Nucl. Phys. {\bf B307}, 864 (1988) ;
            Nucl. Phys. {\bf B310}, 643 (1988) ; I. Klebanov, L. Susskind,
            and T. Banks, Nucl. Phys. {\bf B317}, 665 (1989). 
\item {[4]} S. Giddings and A. Strominger, Nucl. Phys. {\bf B306}, 890
            (1988) ; S. -J. Rey, Nucl. Phys. {\bf B319}, 765 (1989).
\item {[5]} A. Hosoya and W. Ogura, Phys. Lett. {\bf B22}, 117 (1989) ;
            S. -J. Rey, Nucl. Phys. {\bf B336}, 146 (1990).
\item {[6]} K. Lee, Phys. Rev. Lett. {\bf 61}, 263 (1988).

\item {[7]} B. Grinstein and M. B. Wise, Phys. Lett. {\bf B212}, 407 (1988) ; 
            B. Grinstein, Nucl. Phys. {\bf B321}, 439 (1989).
\item {[8]} W. Fischler and L. Susskind, Phys. Lett. {\bf B217}, 48 (1989).

\item {[9]} S. W. Hawking, Commun. Math. Phys. {\bf 43}, 199 (1975).

\item {[10]} S. W. Hawking, Phys. Rev. {\bf D14}, 2460 (1976) ; For a recent
             review, see R. Wald, presented at 1991 Erice lectures 
             (unpublished).
\item {[11]} G. W. Gibbons, S. W. Hawking, and M. J. Perry, Nucl. Phys. 
             {\bf B138}, 141 (1978).
\item {[12]} S. W. Hawking and D. N. Page, Phys. Rev. {\bf D42}, 2655 (1990).

\item {[13]} C. W. Misner, K. S. Thorne, and J. A. Wheeler, {\it Gravitation}
             (W. H. Freeman, San Francisco, 1973).
\item {[14]} J. B. Hartle and S. W. Hawking, Phys. Rev. {\bf D28}, 2960 
             (1983) ; S. W. Hawking, Nucl. Phys. {\bf B239}, 257 (1984).

\item {[15]} A. Vilenkin, Phys. Lett. {\bf B117}, 25 (1982) ; Phys. Rev.
             {\bf D30}, 509 (1984) ; Nucl. Phys. {\bf B252}, 141 (1985) ;
             Phys. Rev. {\bf D33}, 3560 (1986) ; Phys. Rev. {\bf D37},
             888 (1988).

\end{description}

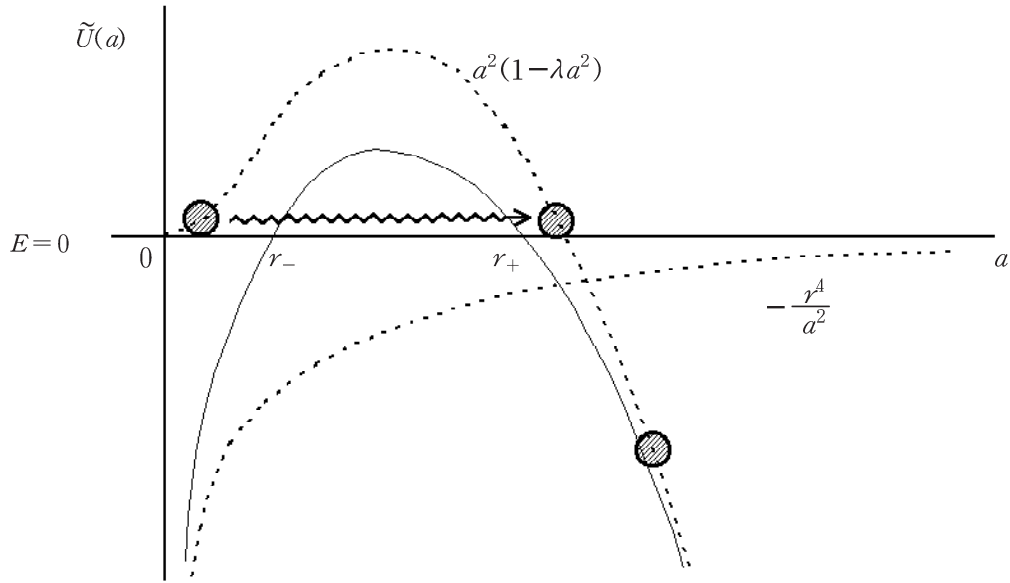
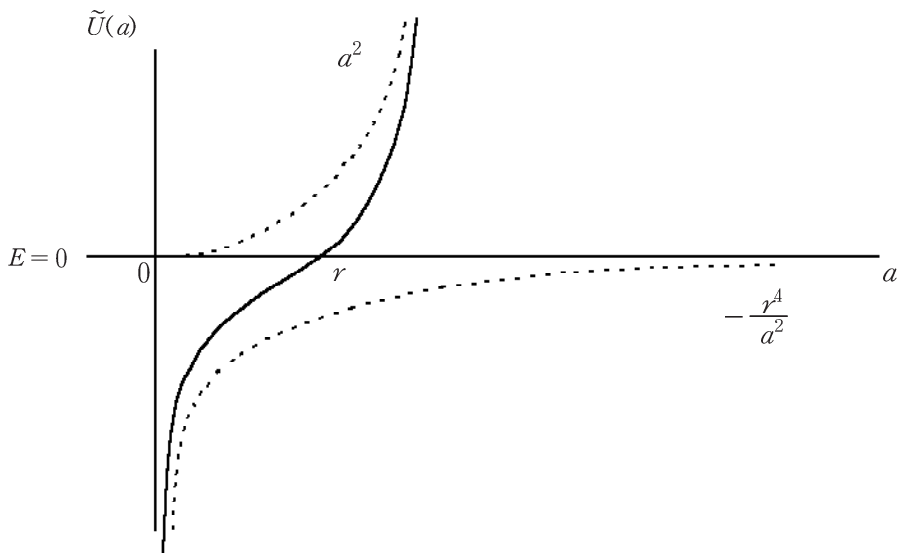
\end{document}